\documentclass[sigconf]{acmart}

\PassOptionsToPackage{protrusion, expansion}{microtype} 

\usepackage{amsmath, amsthm}
    
\usepackage{graphicx, xcolor} 
\usepackage{hyperref, url}
\usepackage{soul}
\usepackage{adjustbox}
\usepackage{array}
\usepackage{booktabs, multirow}
\usepackage{caption}
\usepackage{subcaption}
\usepackage{enumitem}
\usepackage{lipsum}
\usepackage{setspace}
\usepackage{xcolor}

\usepackage{tikz}

\usepackage[ruled, linesnumbered, norelsize]{algorithm2e}

\usepackage{footnote}
\makesavenoteenv{tabular}
\makesavenoteenv{table}
\usepackage[zerostyle=d]{newtxtt}

\setcopyright{rightsretained}

\newcommand{\thiswork}{AMRIC}

\copyrightyear{2023} 
\acmYear{2023} 
\setcopyright{licensedusgovmixed}
\acmConference[SC '23]{The International Conference for High Performance Computing, Networking, Storage, and Analysis}{Nov 12--17, 2023}{Denver, CO}
\acmPrice{15.00}
\acmDOI{XX.XXXX/XXXXXXX.XXXXXXX}
\acmISBN{978-1-4503-8442-1/21/11}

\begin{document}

\title{\thiswork: A Novel In Situ Lossy Compression Framework for Efficient I/O in Adaptive Mesh Refinement Applications}

\settopmatter{authorsperrow=5}

\newcommand{\iu}{Indiana University}
\newcommand{\anl}{Argonne National Lab}
\newcommand{\lanl}{Los Alamos National Lab}
\newcommand{\lbl}{Lawrence Berkeley National Lab}
\newcommand{\pnnl}{Pacific Northwest National Lab}
\newcommand{\fsu}{Florida State University}

\newcommand{\AFFIL}[4]{%
    \affiliation{
        \institution{\small #1}
        \city{#2}\state{#3}\country{#4}
    }
    }

\newcommand{\IU}{\AFFIL{\iu}{Bloomington}{IN}{USA}}
\newcommand{\ANL}{\AFFIL{\anl}{Lemont}{IL}{USA}}
\newcommand{\LANL}{\AFFIL{\lanl}{Los Alamos}{NM}{USA}}
\newcommand{\LBL}{\AFFIL{\lbl}{Berkeley}{CA}{USA}}
\newcommand{\PNNL}{\AFFIL{\pnnl}{Richland}{WA}{USA}}
\newcommand{\FSU}{\AFFIL{\fsu}{Tallahassee}{FL}{USA}}

\author{Daoce Wang}{\IU}
\email{daocwang@iu.edu}

\author{Jesus Pulido}{\LANL}
\email{pulido@lanl.gov}

\author{Pascal Grosset}{\LANL}
\email{pascalgrosset@lanl.gov}

\author{Jiannan Tian}{\IU}
\email{jti1@iu.edu}

\author{Sian Jin}{\IU}
\email{sianjin@iu.edu}

\author{Houjun Tang}{\LBL}
\email{htang4@lbl.gov}

\author{Jean Sexton}{\LBL}
\email{jmsexton@lbl.gov}

\author{Sheng Di}{\ANL}
\email{sdi1@anl.gov}

\author{Zarija Lukić}{\LBL}
\email{zarija@lbl.gov}

\author{Kai Zhao}{\FSU}
\email{ kzhao@cs.fsu.edu}

\author{Bo Fang}{\PNNL}
\email{bo.fang@pnnl.gov}

\author{Franck Cappello}{\ANL}
\email{cappello@mcs.anl.gov}

\author{James Ahrens}{\LANL}
\email{ahrens@lanl.gov}

\author{Dingwen Tao}{\IU}
\authornote{Corresponding author: Dingwen Tao, Department of Intelligent Systems Engineering, Luddy School of Computing, Informatics, and Engineering, Indiana University.}
\email{ditao@iu.edu}
\renewcommand{\shortauthors}{Wang et al.}

\begin{abstract}
As supercomputers advance towards exascale capabilities, computational intensity increases significantly, and the volume of data requiring storage and transmission experiences exponential growth. Adaptive Mesh Refinement (AMR) has emerged as an effective solution to address these two challenges. Concurrently, error-bounded lossy compression is recognized as one of the most efficient approaches to tackle the latter issue. Despite their respective advantages, few attempts have been made to investigate how AMR and error-bounded lossy compression can function together. To this end, this study presents a novel in-situ lossy compression framework that employs the HDF5 filter to improve both I/O costs and boost compression quality for AMR applications. We implement our solution into the AMReX framework and evaluate on two real-world AMR applications, Nyx and WarpX, on the Summit supercomputer. Experiments with 4096 CPU cores demonstrate that AMRIC improves the compression ratio by up to 81$\times$ and the I/O performance by up to 39$\times$ over AMReX's original compression solution.




\end{abstract}



\maketitle

\setlength{\textfloatsep}{6pt}
\setlength\abovecaptionskip{3pt}
\section{Introduction}
\label{sec:introduction}
In recent years, scientific simulations have experienced a dramatic increase in both scale and expense. To address this issue, many high-performance computing (HPC) simulation packages, like AMReX \cite{zhang2019amrex} and Athena++ \cite{stone2020athena++}, have employed Adaptive Mesh Refinement (AMR) as a technique to decrease computational costs while maintaining or even improving the accuracy of simulation results. Unlike traditional uniform mesh methods that apply consistent resolution throughout the entire simulation domain, AMR offers a more efficient approach by dynamically adjusting the resolution and focusing on higher resolution in crucial areas, thereby conserving computational resources and storage needs.

Although AMR data can reduce output data size, the reduction may not be substantial enough for scientific simulations resulting in high I/O and storage costs. For example, an AMR simulation with a resolution of $2048^3$ (i.e., $0.5 \times 1024^3$ mesh points in the coarse level and $0.5 \times 2048^3$ in the fine level) can generate up to 1 TB of data for a single snapshot with all data fields dumped; a total of 1 PB of disk storage is needed, assuming the simulation is run in an ensemble of five times with 200 snapshots dumped per simulation.

To this end, data compression approaches could be utilized in conjunction with AMR techniques to further save I/O and storage costs. However, traditional lossless compression methods have limited effectiveness in reducing the massive amounts of data generated by scientific simulations, typically achieving only up to a compression ratio of 2$\times$. 
As a solution, a new generation of error-bounded lossy compression techniques, such as SZ~\cite{tao2017significantly, di2016fast, sz18}, ZFP~\cite{zfp}, MGARD \cite{ainsworth2018multilevel} and their GPU versions \cite{tian2020cusz,tian2021optimizing, cuZFP}, have been widely used in the scientific community ~\cite{di2016fast,tao2017significantly,zfp,sz18,lu2018understanding,luo2019identifying,tao2019optimizing,cappello2019use,jin2020understanding,grosset2020foresight,jin2022accelerating, baker2019evaluating}. 

While lossy compression holds the potential to significantly reduce I/O and storage costs associated with AMR simulations, there has been limited research on using lossy compression in AMR simulations.
Two recent studies have aimed to devise efficient lossy compression methods for AMR datasets. Luo et al. \cite{zMesh} proposed zMesh, which reorders AMR data across different refinement levels into a 1D array to leverage data redundancy. However, by compressing data in a 1D array, zMesh is unable to exploit higher-dimension compression, leading to a loss of topology information and data locality in higher-dimension data. In contrast, Wang et al. \cite{wang2022tac} developed TAC to enhance zMesh's compression quality through adaptive 3D compression.
While zMesh and TAC offer offline compression solutions for AMR data, they are not suitable for in situ compression of AMR data. We will discuss these works and their limitations for in situ compression in detail in \S\ref{sec:related}.

On the other hand, in situ compression of AMR data could enhance I/O efficiency by compressing data during the application's runtime, allowing for the direct writing of smaller, compressed data to storage systems. This approach would eliminate the need to transfer large amounts of original data between computing nodes and storage systems, further streamlining the process. 
AMReX currently supports in situ compression for AMR data \cite{amrexcomp}; 
however, the current implementation converts the high-dimensional data into a 1D array before compression, which limits the compression performance without the additional spatial information.
Additionally, it utilizes a small HDF5 chunk size, leading to lower compression ratios and reduced I/O performance. These limitations will be discussed in more detail in Sections~\ref{sec:backh5} and~\ref{sec:oph5}.

To address these issues, we propose an effective in situ lossy compression framework for AMR simulations, called \thiswork, that enhances I/O performance and compression quality. Different from AMReX's naïve in situ compression approach, with a customized pre-processing process, \thiswork{} can perform 3D compression and leverage its high compression ratio.
Additionally, we incorporate the HDF5 compression filter to further improve I/O performance and usability. Our primary contributions are outlined below:

\begin{itemize} [topsep=3pt,partopsep=0ex,parsep=0pt]
\item We propose a first-of-its-kind 3D in situ AMR data compression framework through HDF5 (called \thiswork\footnote{The code is available at \url{https://github.com/SC23-AMRIC/SC23-AMRIC}.}).
\item We design a compression-oriented pre-processing workflow for AMR data, which involves removing redundant data, uniformly truncating the remaining data into 3D blocks, and reorganizing the blocks based on different compressors.
\item We employ the state-of-the-art lossy compressor SZ (with two different algorithms/variants) and further optimize it to improve the compression quality for AMR data. This involves utilizing Shared Lossless Encoding (SLE) and adaptive block size in the SZ compressor to enhance prediction quality and hence improve the compression quality.
\item To efficiently utilize the HDF5 compression filter on AMR data, we modify the data layout and the original compression filter to adopt a larger HDF5 chunk size without introducing extra storage overhead. This enables higher compression throughput, improving both I/O and compression quality. 
\item 
We integrate \thiswork{} into the AMReX framework and evaluate it on two real-world AMReX applications, WarpX and Nyx, using the Summit supercomputer. 
\item Experimental results demonstrate that \thiswork{} can significantly outperform non-compression solution and AMReX's original compression solution in terms of I/O performance and compression quality. 
\end{itemize}

The remainder of this paper is organized as follows. In \S\ref{sec:background}, we provide an overview of error-bounded lossy compression for scientific data, the HDF5 file format, AMR approaches, and data structures, as well as a review of related work in the field of AMR data compression. In \S\ref{sec:design}, we describe our proposed 3D in situ compression strategies in detail. 
In \S\ref{sec:evaluation}, we present the experimental results of \thiswork{} and comparisons with existing approaches. 
In \S\ref{sec:related}, we discuss related work and their limitations.
In \S\ref{sec:conclusion}, we conclude this work and outline potential future research.

\section{Background And Motivation}
\label{sec:background}

In this section, we introduce some background information on the HDF5 format and its filter mechanism, lossy compression for scientific data, and AMR methods and AMR data.

\subsection{HDF5 Format and HDF5 Filter}
\label{sec:backh5}
An essential technique for minimizing I/O time associated with the vast data generated by large-scale HPC applications is parallel I/O. Numerous parallel I/O libraries exist, including HDF5 \cite{folk2011overview} and NetCDF \cite{rew1990netcdf}.
In this study, we focus on HDF5, as it is widely embraced by the HPC community and hydrodynamic simulations such as Nyx~\cite{nyx}, a simulation to model astrophysical reacting flows on HPC systems, and VPIC~\cite{bowers2008ultrahigh}, a large-scale plasma physics simulation.
Moreover, HDF5 natively supports data compression filters \cite{hdf5-filter} such as H5Z-SZ~\cite{hdf5filter-sz} and H5Z-ZFP~\cite{hdf5filter-zfp}. 
HDF5 allows chunked data to pass through user-defined compression filters on the way to or from the storage system~\cite{hdf5}. This means that the data can be compressed/decompressed using a compression filter during the read/write operation.

Selecting the optimal chunk size when using compression filters in parallel scenarios is often challenging due to dataset chunking being necessary to enable the use of I/O filters. On the one hand, choosing a chunk size too small may lead to an excessive number of data blocks, resulting in a lower compression ratio caused by the reduction of encoding efficiency and data locality. Additionally, smaller chunks can hamper I/O performance as a consequence of the start-up costs associated with HDF5 and the compressor. On the other hand, larger chunks may enhance writing efficiency but can also create overhead if the chunk size exceeds the data size on certain processors. This occurs because the chunking size must remain consistent across the entire dataset for each process. Striking a balance between these two factors is essential for achieving optimal I/O performance and compression efficiency in parallel environments. Moreover, a larger chunking size can result in the compression of distinct fields of data (e.g., density and velocity) together. To tackle these challenges, we propose an optimized approach to modify the AMR data layout and adaptively determine the maximum chunk size without introducing size overhead issues. We will detail these methods in \S\ref{sec:oph5}.


\subsection{Lossy Compression for Scientific Data}
\label{sec:backlc}

Lossy compression is a common data reduction method that can achieve high compression ratios by sacrificing some non-critical information in the reconstructed data.
Compared to lossless compression, lossy compression often provides much higher compression ratios, especially for continuous floating-point data.
The performance of lossy compression is typically measured by three key metrics: compression ratio, data distortion, and compression throughput.
The compression ratio refers to the ratio between the original data size and the compressed data size.
Data distortion measures the quality of the reconstructed data compared to the original data using metrics such as peak signal-to-noise ratio (PSNR).
Compression throughput represents the size of data that the compressor can compress within a certain time.

In recent years, several high-accuracy lossy compressors for scientific floating-point data have been proposed and developed, such as SZ~\cite{di2016fast, tao2017significantly, sz18} and ZFP~\cite{zfp}.
SZ is a prediction-based lossy compressor, whereas ZFP is a transform-based lossy compressor.
Both SZ and ZFP are specifically designed to compress scientific floating-point data and provide a precise error-controlling scheme based on user requirements.
For example, the error-bounded mode requires users to set a type of error-bound, such as absolute error bound, and a bound value.
The compressor then ensures that the differences between the original and reconstructed data do not exceed the error bound.

In this work, we adopt the SZ lossy compressor in our framework due to its high compression ratio and its modular design that facilitates the integration with our framework, \thiswork{}.
Additionally, the SZ framework includes various algorithms to satisfy different user needs.
For example, SZ with Lorenzo predictor \cite{sz17} provides high compression throughput, while SZ with spline interpolation \cite{sz3} provides high compression ratio, particularly for large error bounds.
Generally, there are three main steps in prediction-based lossy compression, such as SZ.
The first step is to predict each data point's value based on its neighboring points using a best-fit prediction method.
The second step is to quantize the difference between the real value and the predicted value based on the user-set error bound.
Finally, customized Huffman coding and lossless compression are applied to achieve high compression ratios.

Several prior works have studied the impact of lossy compression on reconstructed data quality and post-hoc analysis, providing guidelines on how to set the compression configurations for certain applications~\cite{jin2020understanding,jin2021adaptive,sz3,sz18,sz17,sz16,jin2022accelerating}.
For instance, a comprehensive framework was established to dynamically adjust the best-fit compression configuration for different data partitions of a simulation based on its data characteristics~\cite{jin2021adaptive}.
However, no prior study has established an in situ lossy method for AMR simulations by efficiently leveraging these existing lossy compressors.
Therefore, this paper proposes an efficient in situ data compression framework for AMR simulations to effectively utilize the high compression ratio from lossy compression.

\subsection{AMR Methods and AMR Data}
\label{sec:backamr}

\begin{figure}[t]
     \centering
      \includegraphics[width=0.6\linewidth]{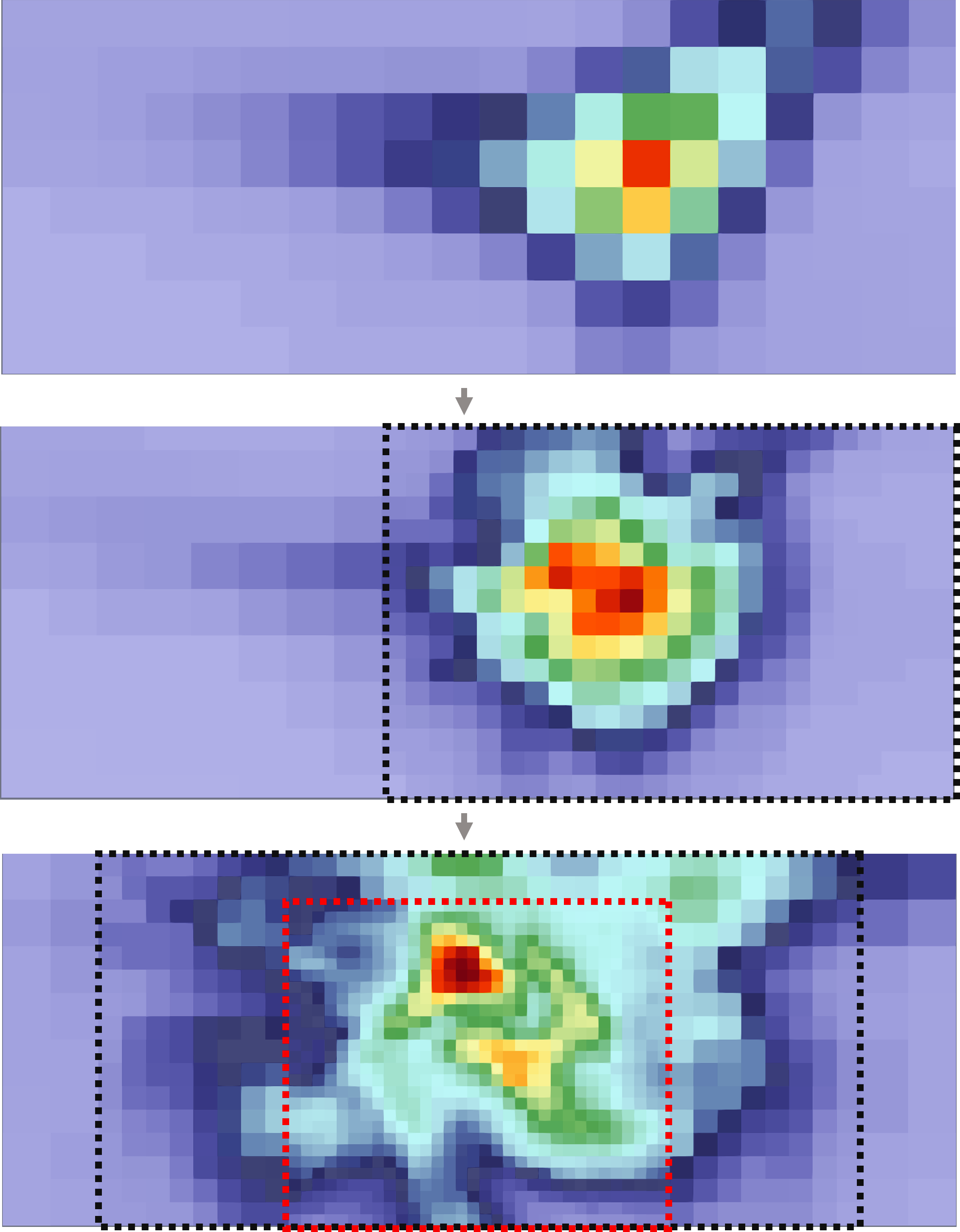}
        \caption[t]{Visualization of an up-close 2D slice of three pivotal timesteps generated by an AMR-based cosmology simulation, Nyx. As the universe evolves, the grid structure adapts accordingly. The dashed black and red boxes highlight areas of finer and finest refinement, respectively.}
        \label{fig:visamr}
        \vspace{-2mm}
\end{figure}

AMR is a technique that tailors the accuracy of a solution by employing a non-uniform grid, which allows for computational and storage savings without compromising the desired accuracy. In AMR applications, the mesh or spatial resolution is adjusted based on the level of refinement required for the simulation. This involves using a finer mesh in regions of greater importance or interest and a coarser mesh in areas of lesser significance. Throughout an AMR simulation, meshes are refined according to specific refinement criteria, such as refining a mesh block when its maximum value surpasses a predetermined threshold (e.g., the average value of the entire field), as illustrated in Figure~\ref{fig:visamr}.

Figure \ref{fig:visamr} shows that during an AMR run, the mesh will be refined when the value meets the refinement criteria, e.g., refining a block when its norm of the gradients or maximum value is larger than a threshold.
By dynamically adapting the mesh resolution in response to the simulation's requirements, AMR effectively balances computational efficiency and solution accuracy, making it a powerful approach for various scientific simulations.

\begin{figure}[t]
    \centering 
    \includegraphics[width=0.98\columnwidth]{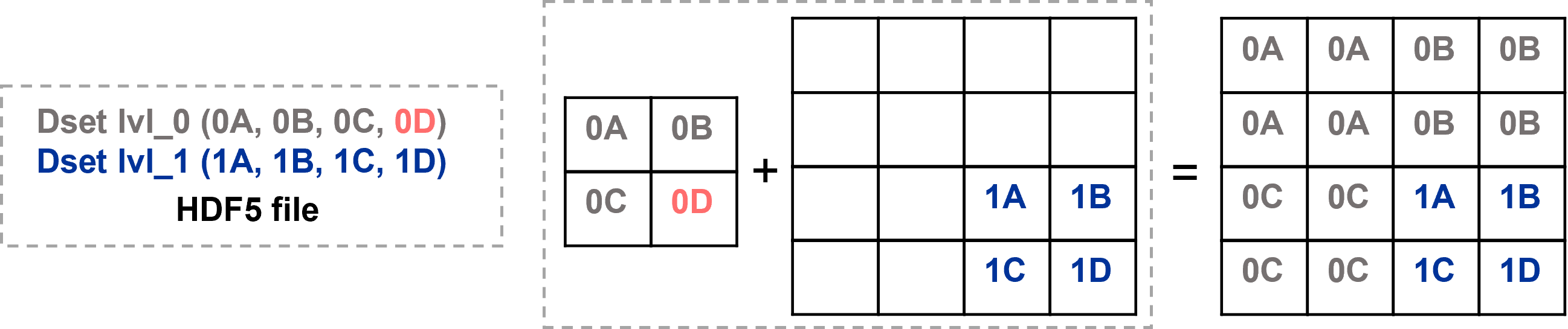}
    \caption{A typical example of AMR data storage and usage.}
   \vspace{-2mm}
    \label{fig:base_ex}
\end{figure}

Data generated by an AMR application is inherently hierarchical, with each AMR level featuring different resolutions. Typically, the data for each level is stored separately, such as in distinct HDF5 datasets(groups). For example, Figure~\ref{fig:base_ex} (left) presents a simple two-level patch-based AMR dataset in an HDF5 structure, where ``0'' indicates the coarse level (low resolution), and ``1'' signifies the fine level (high resolution). When users need the AMR data for post-analysis 
, they usually convert the data from different levels into a uniform resolution. In the given example, the data at the coarse level would be up-sampled and combined with the data at the fine level, excluding the redundant coarse data point ``0D'', as illustrated in Figure~\ref{fig:base_ex} (right). This method could also serve to visualize AMR data without the need for specific AMR visualization tool kits.

\begin{figure*}[h]
\centering
\includegraphics[width=0.88\linewidth]{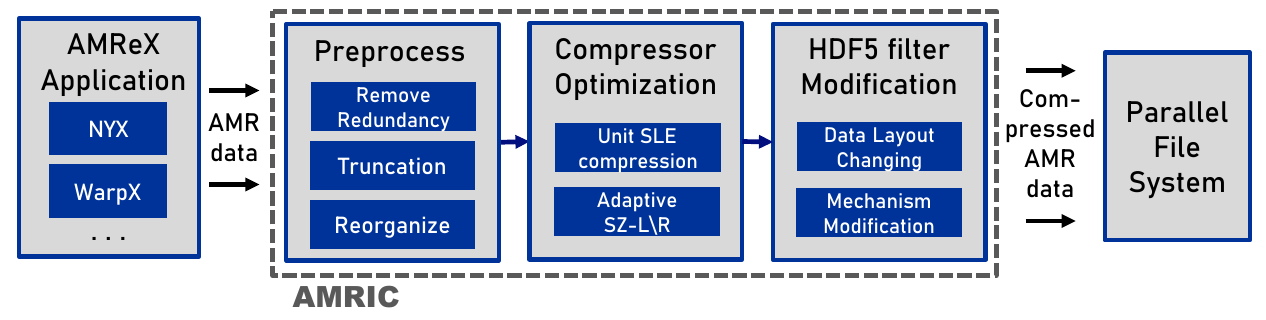}
\vspace{-2mm}
\caption{Overview of our proposed \thiswork.}
\vspace{-4mm}
\label{fig:overview}
\end{figure*}

There are two primary techniques for representing AMR data: patch-based AMR and tree-based AMR~\cite{wang2020cpu}. The key distinction between these approaches lies in how they handle data redundancy across different levels of refinement. Patch-based AMR maintains redundant data in the coarse level, as it stores data blocks to be refined at the next level within the current level, simplifying the computation involved in the refinement process. Conversely, tree-based AMR organizes grids on tree leaves, eliminating redundant data across levels. However, tree-based AMR data can be more complex for post-analysis and visualization when compared to patch-based AMR data~\cite{harel2017two}.

In this work, we focus on the state-of-the-art patch-based AMR framework, AMReX, which supports the HDF5 format and compression filter \cite{amrexcomp}. However, AMReX currently only supports 1D compression, which restricts its ability to leverage higher-dimension compression. Furthermore, the original compression 
of AMReX cannot effectively utilize the HDF5 filter, resulting in low compression quality and I/O performance (will be described in detail in \S\ref{sec:oph5} and Section \ref{sec:related}).
This limitation serves as motivation for our proposal of a 3D in situ compression method for AMReX, aimed at enhancing compression quality and I/O performance.

It is worth noting that the redundant coarser-level data in AMReX (patch-based AMR) is often not utilized during post-analysis and visualization, as demonstrated in Figure~\ref{fig:base_ex} (the coarse point ``0D'' will not be used). Therefore, we discard the redundant data during compression to improve the compression ratio.

\section{Design Methodology}
\label{sec:design}

In this section, we introduce our proposed in situ 3D AMR compression framework, \thiswork{}, using the HDF5 filter, as shown in Figure \ref{fig:overview}, with an outline detailed below.

In \S\ref{sec:de3d}, we first propose a pre-processing approach for AMR data, which includes the elimination of data redundancy, uniform truncation of data, and reorganization of truncated data blocks tailored to the requirements of different compressors including different SZ compression algorithms.
In \S\ref{sec:deco}, we further optimize the SZ compressor's efficiency for compressing AMR data by employing Shared Lossless Encoding (SLE) and dynamically determining the ideal block sizes for the SZ compressor, taking into account the specific characteristics of AMR data.
In \S\ref{sec:oph5}, we present strategies to overcome the obstacles between the HDF5 and AMR applications by modifying the AMR data layout as well as the HDF5 compression filter mechanism, which subsequently results in a significant improvement in both compression ratio and I/O performance, as discussed in \S\ref{sec:backh5}. 

\begin{figure*}[h]
\centering
\includegraphics[width=0.90\linewidth]{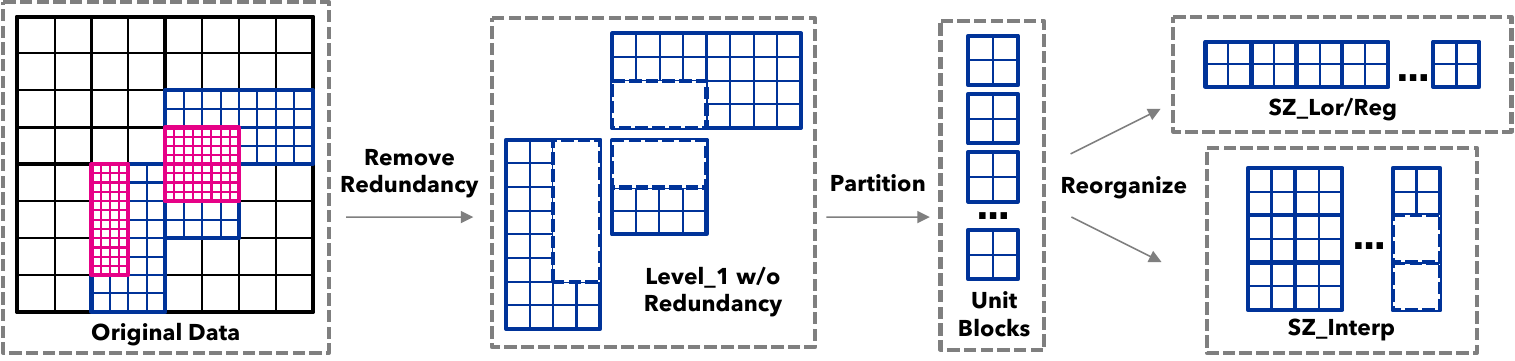}
\caption{An example of our proposed 3D pre-processing workflow (in a top-down 2D view).}
\vspace{-4mm}
\label{fig:3dflow}
\end{figure*}

\subsection{Pre-processing of AMR Data}
\label{sec:de3d}

As mentioned in \S\ref{sec:backamr}, in the patch-based AMR dataset generated by AMReX, the data in the coarse level could be removed to easily improve the compression ratio and I/O performance because there would be fewer data to be processed.
Patch-based AMR divides each AMR level's domain into a set of rectangular boxes.
Figure \ref{fig:3dflow} (right) illustrates an example of an AMR dataset with three total levels.
In the AMReX numbering convention, the coarsest level is designated as level 0. There are 4, 3, and 3 boxes on levels 0, 1, and 2, respectively. Bold lines signify box boundaries.
The four coarsest boxes (black) cover the whole domain.
There are three intermediate-resolution boxes (blue) at level 1 with cells that are two times finer than those at level 0.
The three finest grids (red) at level 2 have cells that are twice as fine as those in level 1.

Clearly, there are overlapping areas between the different AMR levels: the coarsest level 0 overlaps with the finer level 1, and level 1 also has overlapping regions with the finest level 2.
Taking level 1 as an example, we can eliminate the redundant coarse regions that overlap with level 2. It is worth noting that AMReX offers efficient functions for box intersection, which can be employed to identify these overlapping areas. These functions are significantly faster than a naive implementation, resulting in reduced time costs \cite{amr-intersect}. Furthermore, there is no need to record the position of the empty regions in the compressed data, as the position of the empty regions in level 1 can be inferred using the box position of level 2, which introduces minimal overhead to the compressed data size.

The challenge in compressing 3D data is that the boxes will have varying shapes, particularly after redundancy removal, as shown in Figure~\ref{fig:3dflow}, especially in larger datasets. To tackle this irregularity, we propose a uniform truncation method that partitions the data into a collection of unit blocks. This approach facilitates the collective compression of boxes, irrespective of their varying and irregular shapes. This method not only boosts encoding efficiency but also reduces the compressor's launch time by eliminating the need to call the compressor separately for each unique box shape.


Subsequently, the generated unit blocks after truncation, as mentioned, can be rearranged based on the needs of a specific compressor to improve compression performance. In this work, we focus on the SZ2 compression algorithm with the Lorenzo and linear regression predictors (denoted by ``SZ\_L/R'') \cite{sz18}, and the SZ compression algorithm with the spline interpolation approach (denoted by ``SZ\_Interp'') \cite{zhao2021optimizing}.
Specifically, the SZ\_L/R will first truncate the whole input data to blocks with the size of 6$\times$6$\times$6, and then perform Lorenzo predictor or high-dimensional linear regression on each block separately.
For the SZ\_L/R, we will linearize the truncated unit blocks as shown in the top right of Figure \ref{fig:3dflow} (put the unit blocks along the z-axis for 3D data) because this will offer the minimum operation during the organization, thus saving time.

\begin{figure}[t]
     \centering
     \begin{subfigure}[t]{0.95\columnwidth}
         \centering
         \includegraphics[width=\linewidth]{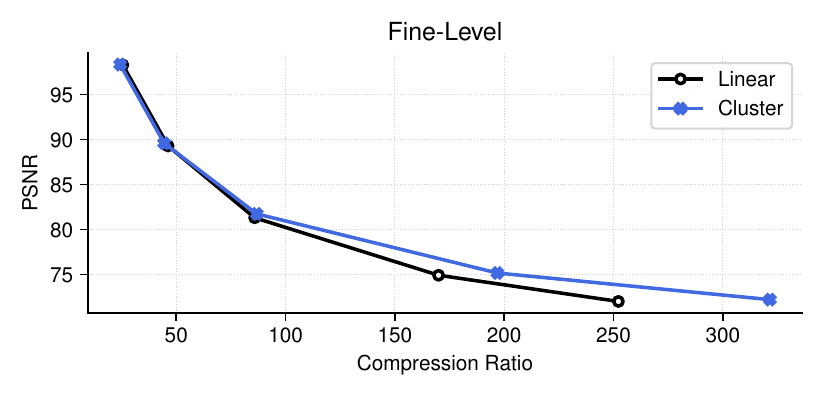}
         \vspace{-6mm}
         \caption{Fine level (density = 17.4\%)}
         \label{fig:sz3-fine}
     \end{subfigure}
     \begin{subfigure}[t]{0.95\columnwidth}
         \centering
         \includegraphics[width=\linewidth]{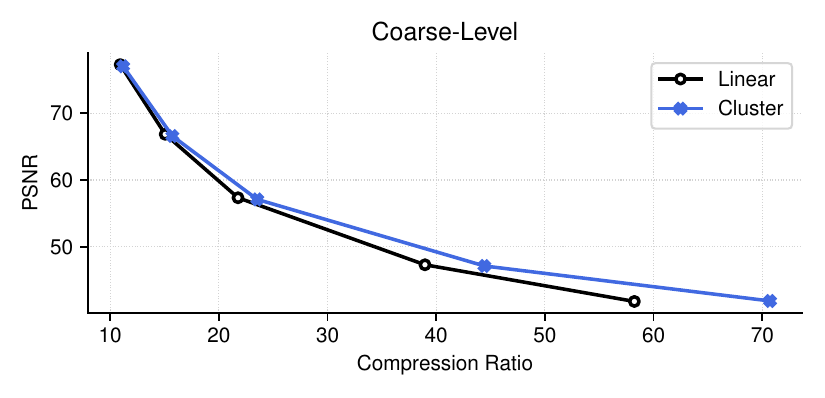}
         \vspace{-6mm}
         \caption{Coarse level (density = 82.3\%)}
         \label{fig:sz3-coarse}
     \end{subfigure}
        \caption[t]{Rate-distortion comparison between linear and cluster arrangements across different levels for Nyx's "baryon density" field. The considered relative error bounds range from $2 \times 10^{-2}$ to $3 \times 10^{-4}$.
        }
        \vspace{-2mm}
        \label{fig:sz3}
\end{figure}

The SZ\_Interp will perform interpolation across all three dimensions of the entire dataset. Given that interpolation is a global operation, one potential solution to improve interpolation accuracy is to cluster the truncated unit blocks more closely into a cube-like formation, as depicted in the bottom right part of Figure \ref{fig:3dflow}. This configuration helps balance the interpolation process across multiple dimensions, thus significantly improving the compression performance.
As depicted in Figure \ref{fig:sz3}, organizing unit blocks in a more compact cluster arrangement leads to enhanced overall compression performance concerning rate-distortion (PSNR\footnote{PSNR is calculated as $20\cdot \log_{10} R - 10\cdot \log_{10}\left(\sum_{i=1}^{N} {e_i^2}/{N}\right)$, where $e_i$ is the absolute error for the point $i$, $N$ is the number of points, and $R$ is the value range of the dataset.} versus compression ratio) when compared to a linear arrangement of the blocks. This improvement is particularly noticeable when the compression ratio is relatively high.
The test data is generated from a Nyx Run featuring two refinement levels: a coarse level with $256^3$ grids and a fine level containing $512^3$ grids. After eliminating redundant coarse data, the coarse level has a data density of 82.3\%, while the fine level has a data density of 17.4\%. Here, data density refers to the proportion of data saturation within the entire domain.

\subsection{Optimization of SZ\_L/R Compression}
\label{sec:deco}
The pre-processed AMR data, however, faces two significant challenges that prevent it from achieving optimized compression quality when using the original SZ\_L/R compressor. To overcome these challenges and further improve compression performance, we propose optimizing the SZ\_L/R compressor. In the following paragraphs, we will outline the two challenges and describe our proposed solutions to address them effectively.

\begin{figure}[t]
         \centering
         \includegraphics[width=\linewidth, ]{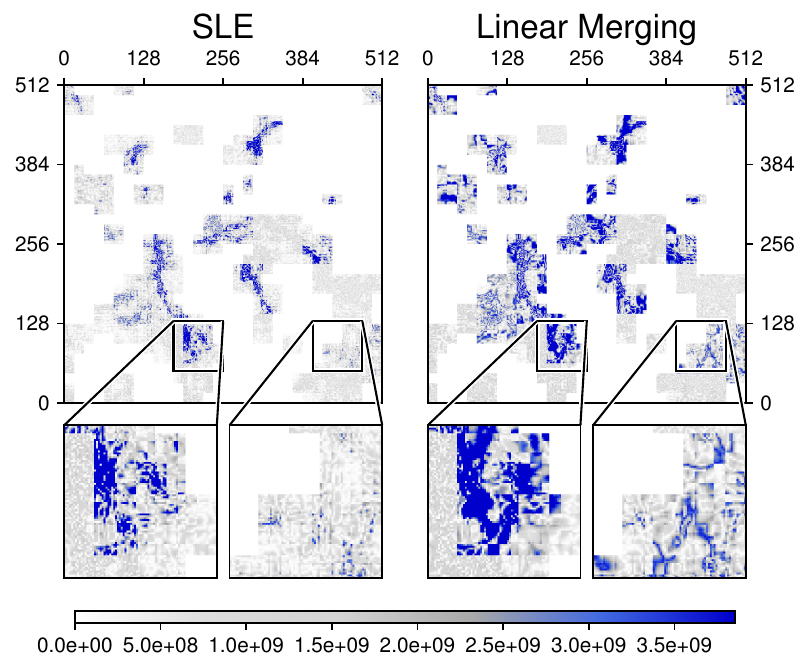}
        \caption[t]{Visualization comparison (one slice) of absolute compression errors of unit SLE (left, CR = 91.4) and original linear merging (right, CR = 86.1) on Nyx “baryon density” field (i.e., fine level, 18\% density, unit block size~=~16). Bluer means higher compression error.} 
        \vspace{-2mm}
        \label{fig:slevis}
\end{figure}

\textit{\textbf{Challenge 1}: Low prediction accuracy of SZ\_L/R on AMR data.} As discussed in \S\ref{sec:de3d}, the truncated unit blocks are linearized and sent to the SZ\_L/R compressor. However, some merged small blocks may not be adjacent in the original dataset, resulting in poor data locality/smoothness between these non-neighboring blocks. This negatively affects the accuracy of SZ\_L/R's predictor.
An intuitive solution would be to compress each box individually. But, truncation can produce a large number of small data blocks (e.g., 5,000+), causing the SZ\_L/R to struggle on small datasets due to low encoding efficiency, as mentioned in \S\ref{sec:backh5}. This is because the SZ compressor utilizes thousands of Huffman trees to encode these small blocks separately, leading to decreased encoding efficiency.
In conclusion, the original SZ\_L/R faces a dilemma: either predict and encode small blocks collectively (by merging them), which compromises prediction accuracy, or predict and encode each small block individually, incurring high Huffman encoding overhead.

\textit{\textbf{Solution 1}: Improve prediction using unit SLE.} To address Challenge 1, we propose using the Shared Lossless Encoding (SLE) technique in SZ\_L/R. This method allows for separate prediction of unit data blocks while encoding them together with a single shared Huffman tree.
Specifically, each unit block is initially predicted and quantized individually. Afterward, the quantization codes and regression coefficients from each unit block are combined to create a shared Huffman tree and then encoded. This approach improves the prediction performance of SZ\_L/R without significantly increasing the time overhead during the encoding process.

\begin{figure}[t]
     \centering
     \begin{subfigure}[t]{\columnwidth}
         \centering
         \includegraphics[width=\linewidth]{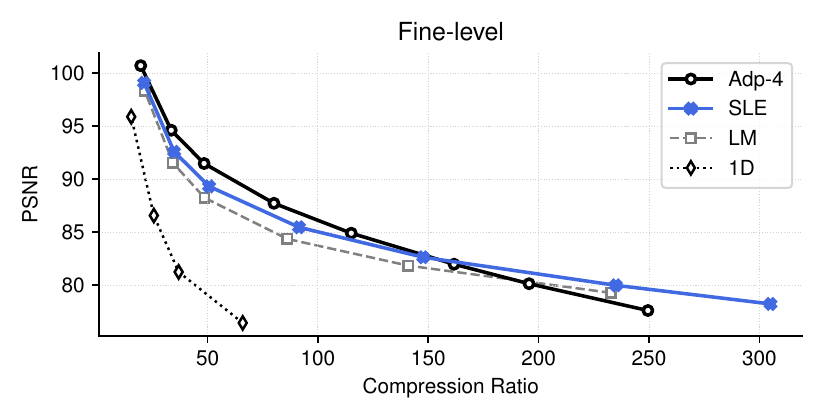}
         \vspace{-6mm}
         \caption{Fine level (unit block size=16, density = 17.4\%)}
         \label{fig:sz2-fine}
     \end{subfigure}
     \begin{subfigure}[t]{\columnwidth}
         \centering
         \includegraphics[width=\linewidth]{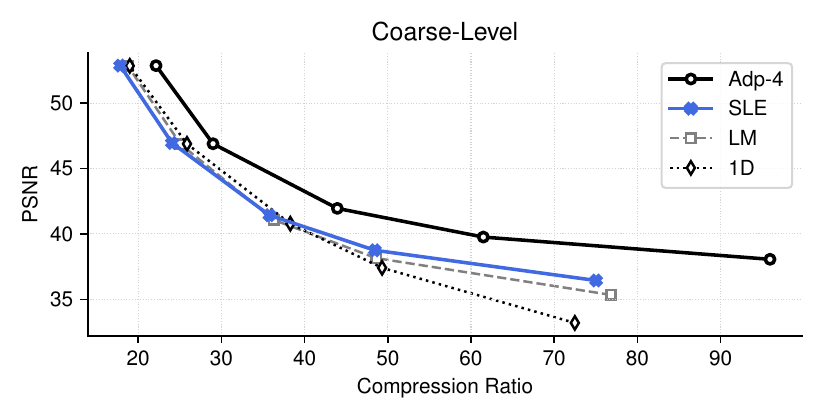}
         \vspace{-6mm}
         \caption{Coarse level (unit block size=8, density = 82.3\%)}
         \label{fig:sz2-coarse}
     \end{subfigure}
        \caption[t]{Rate-distortion comparison between LM, SLE, adaptive SZ\_L/R, and 1D compression, across different levels for Nyx's "baryon density" field. The relative error bound ranges from $2 \times 10^{-2}$ to $3 \times 10^{-4}$.
        }
        \vspace{-2mm}
        \label{fig:sz2}
\end{figure}

As shown in Figure~\ref{fig:slevis}, the unit SLE notably reduces overall compression error in comparison to the original linear merging (LM), especially for data located at the boundaries of data blocks. As a result, this leads to a substantial improvement in rate distortion, as illustrated in Figure~\ref{fig:sz2-fine}.
Note that the data used for testing in this section is the same as in \S\ref{sec:de3d}. Specifically, the unit block size for the fine level is 16, while the unit block size for the coarse level is 8.


\begin{figure}[h]
     \centering
     \begin{subfigure}[t]{0.33\linewidth}
         \centering
         \includegraphics[width=\linewidth]{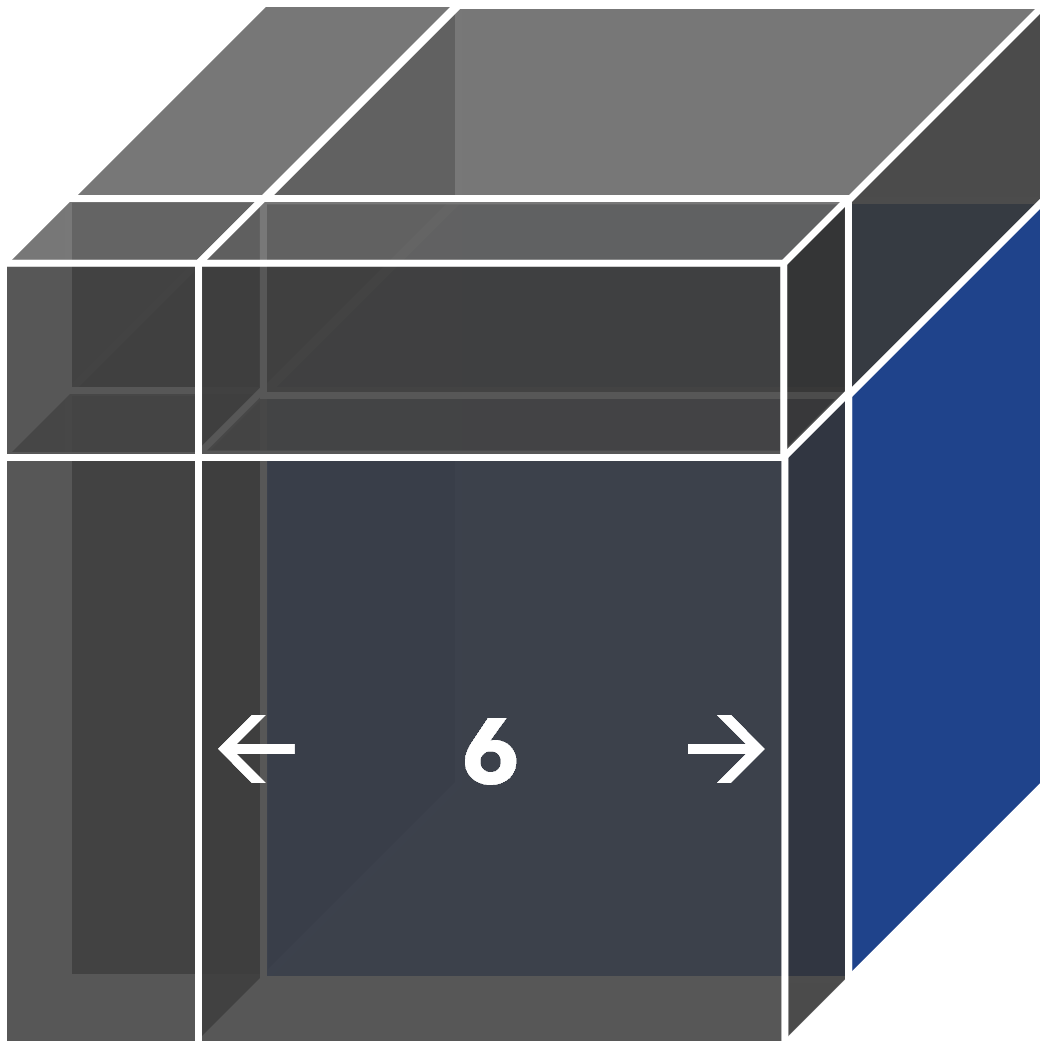}  
         \caption[t]{Original partition}
         \label{fig:sz6}
     \end{subfigure}\hspace{0.1\linewidth}
     \begin{subfigure}[t]{0.33\linewidth}
         \centering
         \includegraphics[width=\linewidth]{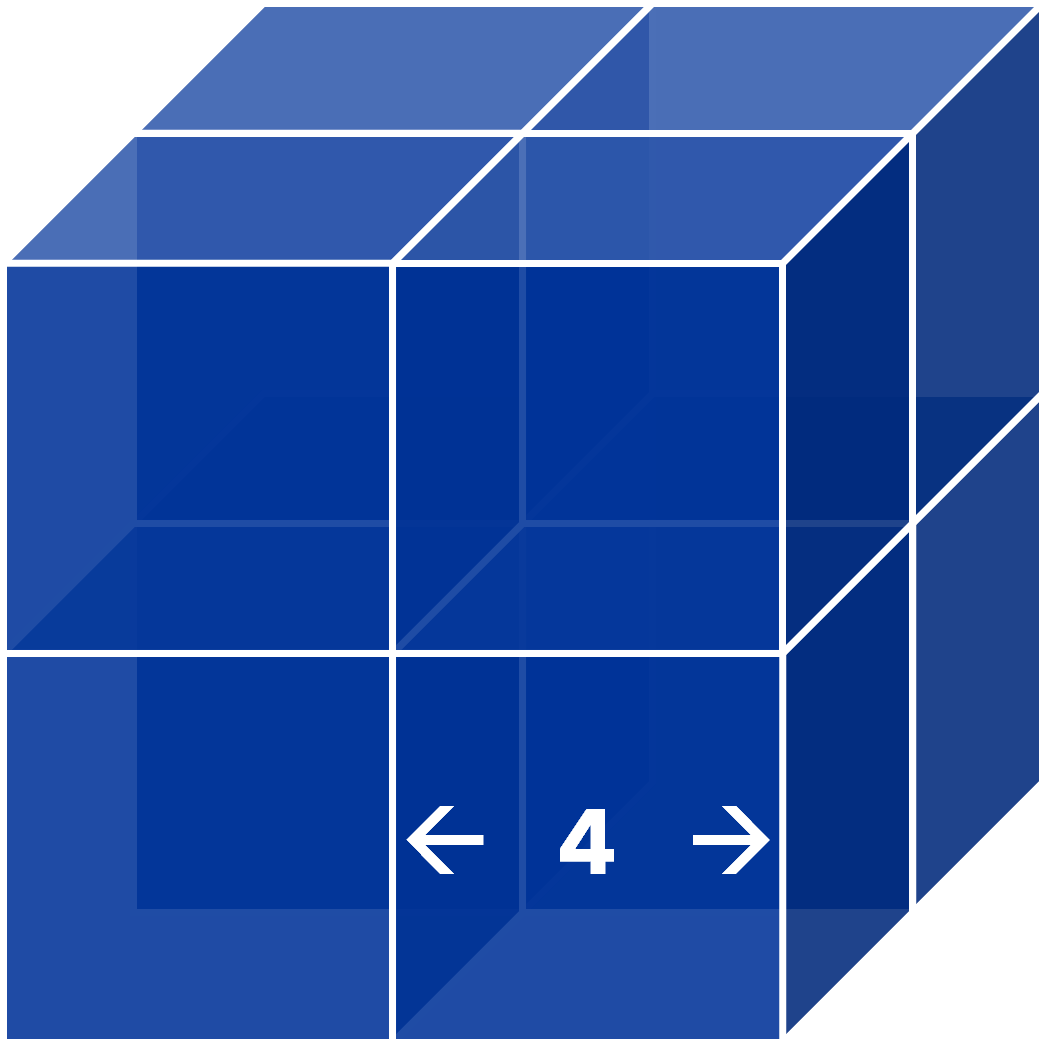}
         \caption{Adaptive partition}
         \label{fig:sz4}
     \end{subfigure}
        \caption[t]{Example of the original partition and adaptive partition of SZ\_L/R on a unit block with the size of 8$\times$8$\times$8; the gray boxes represent data that are difficult to compress.}
        \label{fig:adpsz}
        \vspace{-2mm}
\end{figure}

\textit{\textbf{Challenge 2}: Unit SLE may produce undesirable residues.}
As previously mentioned, the input data is truncated into 6$\times$6$\times$6 blocks by the SZ\_L/R compressor for separate processing. This block size was chosen to balance between prediction accuracy and metadata overhead, achieving optimal overall compression quality.

\begin{figure}[b]
     \centering
         \includegraphics[width=\linewidth]{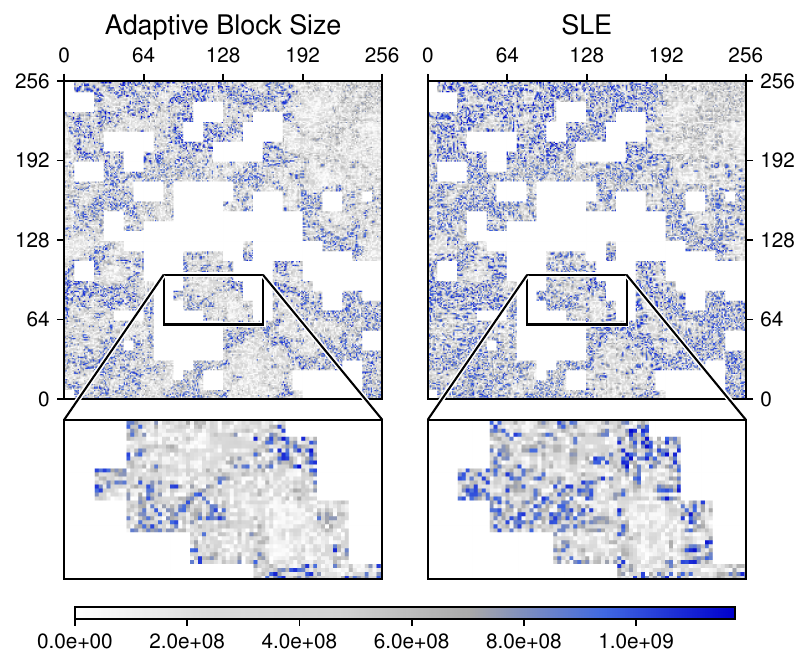}
         \vspace{-6mm}
        \caption[t]{Visualization comparison (one slice) of compression errors of the adaptive block size (left, CR=39.8) and unit SLE (right, CR=38.8) and  on Nyx ``baryon density'' field (i.e., coarse level, 82\% density, unit block size~=~8). Bluer means higher compression error.} 
        \label{fig:adpszvis}
\end{figure}

\begin{figure*}[]
     \centering
     \begin{subfigure}[t]{0.31\linewidth}
         \centering
         \includegraphics[width=\linewidth]{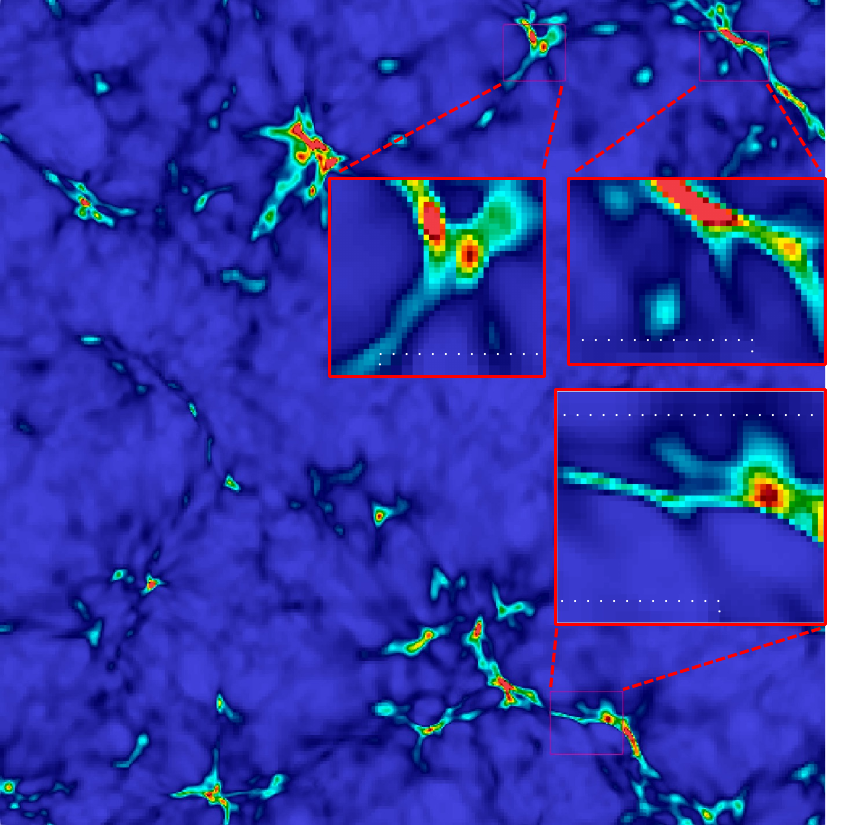}  
         \caption[t]{Original data}
         \label{fig:all-ori}
     \end{subfigure} 
     \begin{subfigure}[t]{0.31\linewidth}
         \centering
         \includegraphics[width=\linewidth]{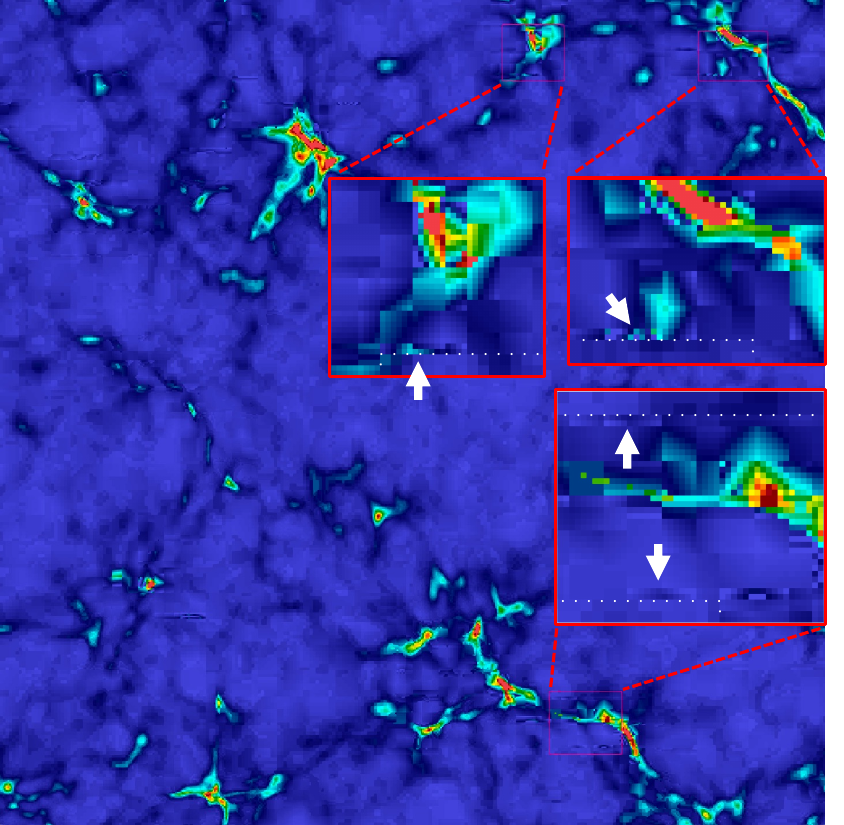}
         \caption{Original SZ\_L/R, CR~=51.7}
         \label{fig:all-sz}
     \end{subfigure} 
     \begin{subfigure}[t]{0.31\linewidth}
         \centering
         \includegraphics[width=\linewidth]{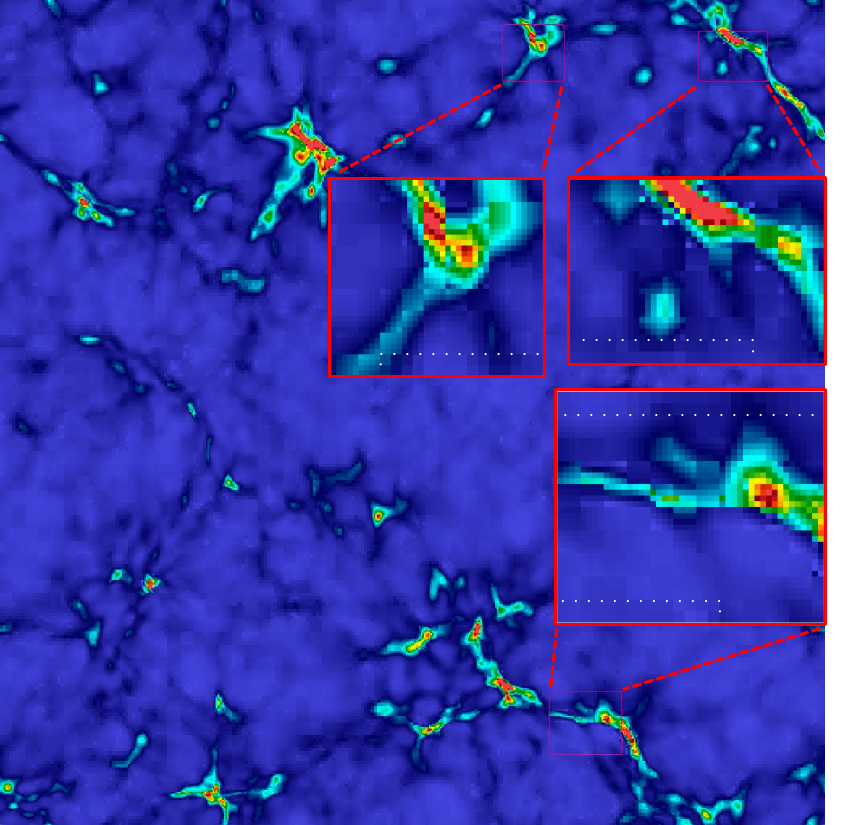}
         \caption{AMRIC SZ\_L/R, CR~=53.2}
         \label{fig:all-amric}
     \end{subfigure}
        \caption[t]{
        Visualization comparison (one slice) of original (uncompressed) data and decompressed data produced by original SZ\_L/R and AMRIC's optimized SZ\_L/R on Nyx's ``baryon density'' field (with 2 levels, 18\% density for fine level). Warmer colors indicate higher values. The red dotted lines denote the boundaries between AMR levels, and the white arrows in Figure~\ref{fig:all-sz} highlight the artifacts between two AMR levels.}
        \vspace{-4mm}
        \label{fig:all}
\end{figure*}

When using unit SLE, the compressor will further partition each of these unit blocks. The issue is that the unit block size of data produced by AMReX is typically a power of two (i.e., $2^n$), which is not evenly divisible by 6. As a result, using a 6$\times$6$\times$6 cube to truncate unit blocks with specific sizes may leave undesirable residues that impact compression quality.
For example, if the unit block is 8$\times$8$\times$8, as shown in Figure~\ref{fig:adpsz}, SZ\_L/R with unit SLE will further divide it into smaller blocks with sizes of 6$\times$6$\times$6 (one block), 6$\times$6$\times$2 (three ``flat'' blocks), 6$\times$2$\times$2 (three ``slim'' blocks), and 2$\times$2$\times$2 (one ``tiny'' block) as shown in Figure~\ref{fig:sz6}. While the data in the 6$\times$6$\times$2, 6$\times$2$\times$2, and 2$\times$2$\times$2 blocks is almost flattened/collapsed to 2D data, 1D data, and a single point, respectively, rather than preserving 3D data features. These ``low-dimension" data blocks can greatly affect the prediction accuracy of SZ\_L/R, as they cannot leverage high-dimensional topological information. As shown in Figure~\ref{fig:sz2-coarse}, when the unit block size is 8, the unit SLE approach does not appear to significantly improve the performance over the original LM method.

\textit{\textbf{Solution 2}: SZ\_L/R with adaptive block size.} 
To address the issue of residue blocks that are difficult to compress, we propose an adaptive approach for selecting the block size used by the SZ\_L/R compressor based on the unit block size of the AMR data. Equation~\ref{eq:1} describes the adaptive block size selection method.
\begin{equation}
    \text{SZ\_BlkSize} =
    \begin{cases}
      $4$\times$4$\times$4$, & \text{if}\ \text{unitBlkSize} \mod 6 \le 2; \\
      $6$\times$6$\times$6$, & \text{if}\ \text{unitBlkSize} \mod 6 > 2; \\
      $6$\times$6$\times$6$, & \text{if}\  \text{unitBlkSize} \ge 64;
    \end{cases}
\label{eq:1}
\end{equation}

Specifically, if the remainder of the unit block size divided by the original SZ\_L/R block size is less than or equal to 2, there will be undesirable residue blocks. In such cases, we adjust the SZ\_L/R block size to be $4\times 4\times 4$ to avoid compressing these blocks with low compressibility and to improve prediction quality. Conversely, if the remainder is greater than 2, we use the original block size of $6\times 6\times 6$.
For example, as shown in Figure~\ref{fig:sz4}, for the 8$\times$8$\times$8 unit block, we have $8~mod~6=2$, and we will select the SZ\_L/R block to be $4^3$.
Although using an SZ\_L/R block size of $4^3$ results in higher metadata overhead, the increased prediction accuracy compensates for it, achieving compression performance comparable to that of $6^3$ while avoiding the undesirable residue issue.
Figure~\ref{fig:adpszvis} illustrates that the adaptive block size approach (i.e., Adp-4) can significantly reduce compression errors compared to the SLE approach, leading to a considerable enhancement in rate-distortion, as shown in Figure~\ref{fig:sz2-coarse}.
On the other hand, for example, when the unit block size is 16, we do not have the undesirable residues issue and there is no need to use an adaptive block size approach since it does not have an obvious advantage over the SLE approach, as shown in Figure~\ref{fig:sz2-fine}.
Furthermore, note that when the unit block size is relatively large (i.e., larger than 64, which is not common for AMR data), we retain the original SZ\_L/R block size. This is because even if there are undesirable residue blocks, they only occupy a small portion of the dataset, while the compression performance using $6^3$ is slightly better than using $4^3$, offsetting the negative effect of the few residue blocks.
Note that we did not select the SZ\_L/R block size to be $8^3$ to eliminate the undesirable residue because it will significantly reduce compression quality.

\textit{\textbf{Improvement in Visualization Quality}}: It is worth noting that compared to the original SZ\_L/R, AMRIC's optimized SZ\_L/R notably enhances the visualization quality of AMR data, particularly in areas with intense data fluctuations. For example, as shown in Figure~\ref{fig:all}, 
our optimized SZ\_L/R effectively reduces the artifacts at the boundaries between different AMR levels (denoted by the white dotted lines), which were previously caused by the original SZ\_L/R (as indicated by the white arrows in Figure~\ref{fig:all-sz}).


\subsection{Modification of HDF5 Compression Filter}
\label{sec:oph5}
In this work, we use the HDF5 compression filter to enhance I/O performance and increase usability. However, as mentioned in \S\ref{sec:backh5}, there are barriers between the HDF5 filter and the AMR application.
Specifically, when employing the compression filter, HDF5 splits the in-memory data into multiple chunks, with filters being applied to each chunk individually. This makes it challenging to determine an appropriate large chunk size in order to improve the compression ratio and I/O performance. We face two primary obstacles when attempting to use a larger chunk size, and we will discuss each of these challenges along with their solutions.

\begin{figure}[t]
\centering
 \includegraphics[width=0.85\linewidth]{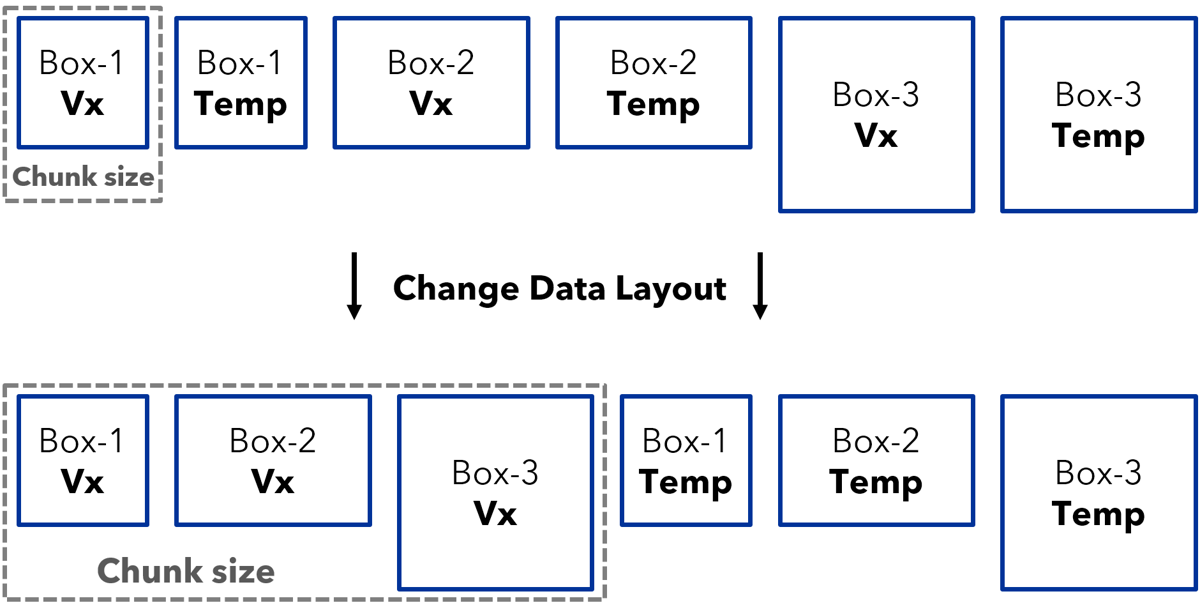}
\caption{An example of an AMR dataset with two fields and three boxes, illustrating our data layout modification that applies a larger chunk size (indicated by the grey dashed line).}
\label{fig:layout}
\vspace{-2mm}
\end{figure}

\textit{\textbf{Challenge 1}: AMR data layout issue for multiple fields.} As discussed in \S\ref{sec:de3d}, AMReX (Patch-based AMR) divides each AMR level's domain into a collection of rectangular boxes/patches, with each box typically containing data from multiple fields.
Consequently, in the AMReX framework, 
data corresponding to various fields within each box are stored continuously, rather than being stored in a separate manner.
For instance, as illustrated in the upper portion of Figure~\ref{fig:layout}, we have three boxes and two fields (i.e., Temp for temperature and Vx for velocities in the $x$ direction), and the data for Temp and Vx in different boxes are placed together. In this situation, when determining the HDF5 chunk size, it cannot exceed the size of the smallest box (i.e., Box-1).

This limitation arises because we want to avoid compressing different fields together using lossy compression, as it would lead to the usage of identical error bounds across various fields, despite their potentially significant differences in value ranges.
Additionally, combining data from distinct fields can compromise data smoothness and negatively impact the compressor's performance.

As a result, the original AMReX could only utilize a small chunk size (i.e., 1024), which significantly increased the encoding overhead of compressing each small chunk separately and led to a lower compression ratio. Moreover, the compressor had to be called for each small chunk, substantially raising the overall startup cost of the compressor and adversely affecting I/O performance.

\textit{\textbf{Solution 1}: Change data layout.} A potential solution to this issue is to separate data from different fields into distinct buffers for compression. However, this approach requires compressing and writing multiple buffers into multiple HDF5 datasets simultaneously, resulting in reduced performance for HDF5 collective writes. Based on our observations, compressing and writing AMR data into multiple HDF5 datasets can be up to 5$\times$ slower than processing them collectively.

To address this problem, we propose continuing to compress and write data into a single HDF5 dataset, while modifying the data layout to group data from the same field of each box together, as depicted in the lower portion of Figure~\ref{fig:layout}. This method allows us to increase the chunk size and compress the entire field as a whole. It is important to note that we achieved this by altering the loop access order when reading the data into the buffer, which adds minimal time overhead, rather than reorganizing the buffer itself. By increasing the chunk size, we can significantly enhance both the compression and I/O performance (will be shown in \S\ref{sec:evaluation}).

\textit{\textbf{Challenge 2}: Load imbalance for AMR data.}
Another challenge that prevents the adoption of a large chunk size is the load imbalance issue for AMR data across multiple processes.
Given that the entire HDF5 dataset has to use the same chunk size, selecting an optimal global chunk size in a parallel scenario becomes difficult, as the data size on each MPI rank may vary.

\begin{figure}[t]
\centering
 \includegraphics[width=0.75\linewidth]{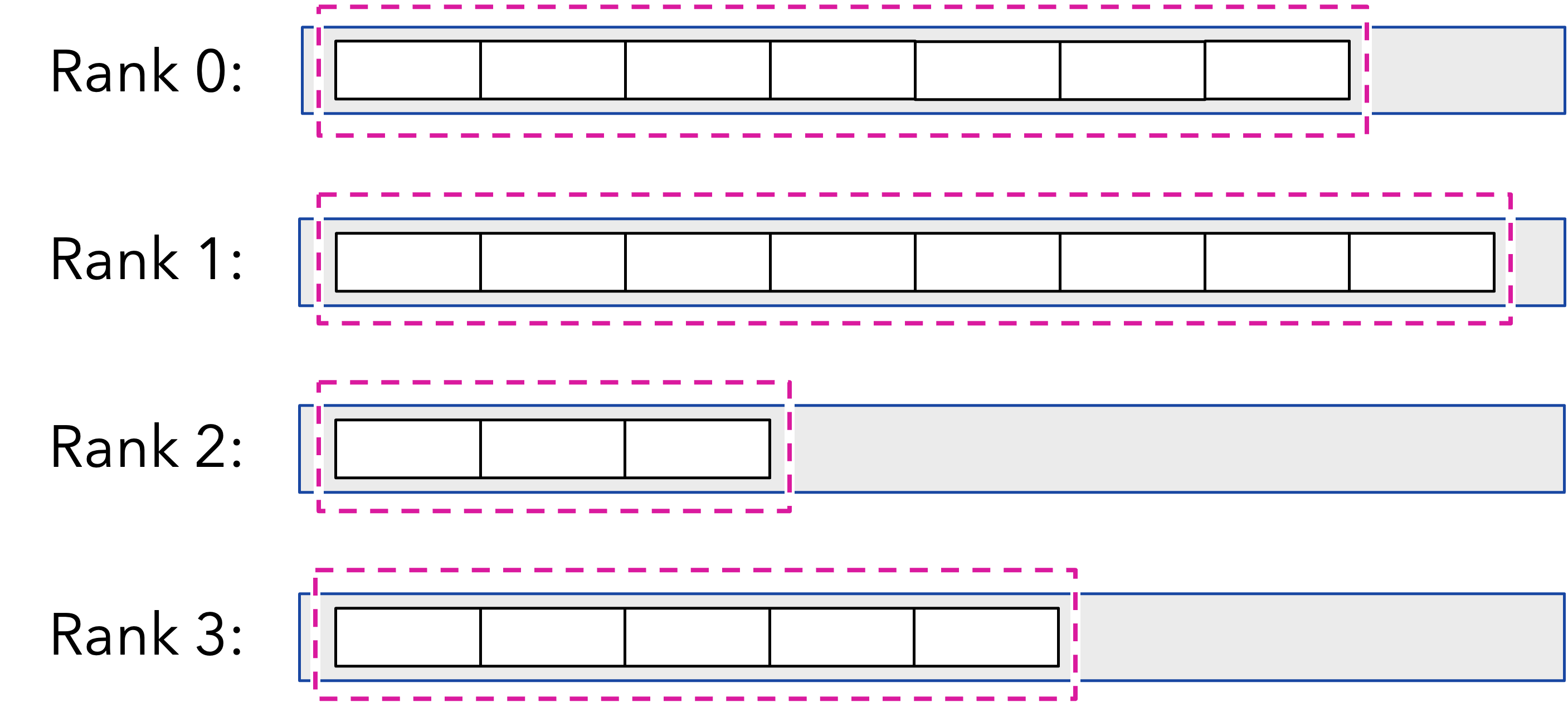}
\caption{An example with four MPI ranks that hold a different amount of AMR data to demonstrate our proposed chunk size selection strategy. We select the chunk size to be the data size of rank 1 (outer blue box) while passing the actual data size to the compression filter (enclosing magenta dashed box).}
\label{fig:fill}
\vspace{-2mm}
\end{figure}

For example, as shown in Figure~\ref{fig:fill}, we have 4 ranks that hold different data in the memory. Without loss of generality and for clearer demonstration, we suppose there is only one field.
If we set the chunk size to be the largest data size in all the ranks (i.e., rank 1), there would be overhead in the other 3 ranks (i.e., we have to pad useless data onto the other 3 ranks). Clearly, This will make the compressor handle extra data and impact the compression ratio as well as the I/O time. 

Another intuitive solution to this is to let each rank write its data to its own dataset. In this way, each rank does not have to use the same global chunk size and can select its own chunk size based on the data size. The problem is that due to the usage of the filter, HDF5 has to perform the collective write, which means all processes need to participate in creating and writing each dataset. For example, when rank 0 is writing its data to dataset 0, the other 3 ranks also need to participate even if they have no data to be written to dataset 0. As a result, the other 3 ranks will be idle and wait for the current write (to dataset 0) to finish before proceeding with its own write, causing a serial write with poor performance.

\textit{\textbf{Solution 2}: Modify the HDF5 filter mechanism.}
To address the above issue, we propose to still use the global chunk size, which is equal to the largest data size across all ranks. However, we modify the compression filter and provide the actual data size of each rank (as shown in the magenta dashed box in Figure~\ref{fig:fill}) to the compression filter before the compression process.

It should be noted that we also need to store metadata such as the value of the original data size for each rank for decompression purposes. The metadata overhead is minimal since the data component far outweighs the metadata to be written.
This results in nearly no size overhead while improving the overall compression ratio and I/O time by adapting the biggest possible chunk size.

\section{Experimental Evaluation}
\label{sec:evaluation}
\begin{table*}[h]
\caption{Detailed information about our tested AMR runs.}
\resizebox{.9\linewidth}{!}{
\begin{tabular}{|c|c|c|c|c|c|c|}
\hline
\textbf{Runs} &
  \textbf{\#AMR Levels} &
  \begin{tabular}[c]{@{}c@{}}\textbf{\#Nodes} \\ (\#MPI ranks)\end{tabular}  &
  \begin{tabular}[c]{@{}c@{}}\textbf{Grid size of each level} \\ (coarse to fine)\end{tabular} &
  \begin{tabular}[c]{@{}c@{}}\textbf{Density of each level} \\ (coarse to fine)\end{tabular} &
  \begin{tabular}[c]{@{}c@{}}\textbf{Data size} \\ (each timestep)\end{tabular} & 
  \begin{tabular}[c]{@{}c@{}}\textbf{Error bound} \\ (AMRIC and  AMReX)\end{tabular} \\ \hline
Warpx\_1 & 2 & 2 (64)   & 256$\times$256$\times$2048, 512$\times$512$\times$4096   & 98.04\%, 1.96\% & 12.4 GB & 1E-3, 5E-3 \\ \hline
Warpx\_2 & 2 & 16 (512) & 512$\times$512$\times$4096, 1024$\times$1024$\times$8192 & 98.05\%, 1.96\% & 99.3 GB & 1E-3, 5E-3 \\ \hline
Warpx\_3 & 2 & 128 (4096) & 1024$\times$1024$\times$8192, 2048$\times$2048$\times$16384 & 98.96\%, 1.04\% & 624 GB & 1E-4, 5E-4 \\ \hline
Nyx\_1   & 2 & 2 (64)   & 256$\times$256$\times$256, 512$\times$512$\times$512     & 98.6\%, 1.4\%   & 1.6 GB & 1E-3, 1E-2 \\ \hline
Nyx\_2   & 2 & 16 (512) & 512$\times$512$\times$512, 1024$\times$1024$\times$1024  & 96.67\%, 3.23\% & 12 GB & 1E-3, 1E-2  \\ \hline
Nyx\_3   & 2 & 128 (4096) & 1024$\times$1024$\times$1024, 2048$\times$2048$\times$2048  & 98.3\%, 1.7\% & 97.5 GB & 1E-3, 1E-2  \\ \hline
\end{tabular}
}
\vspace{-2mm}
\label{tab:dataset}
\end{table*}

\subsection{Experimental Setup}
\begin{figure}[t]
     \centering
         \centering
         \vspace{-2mm}
         \includegraphics[width=0.7\columnwidth]{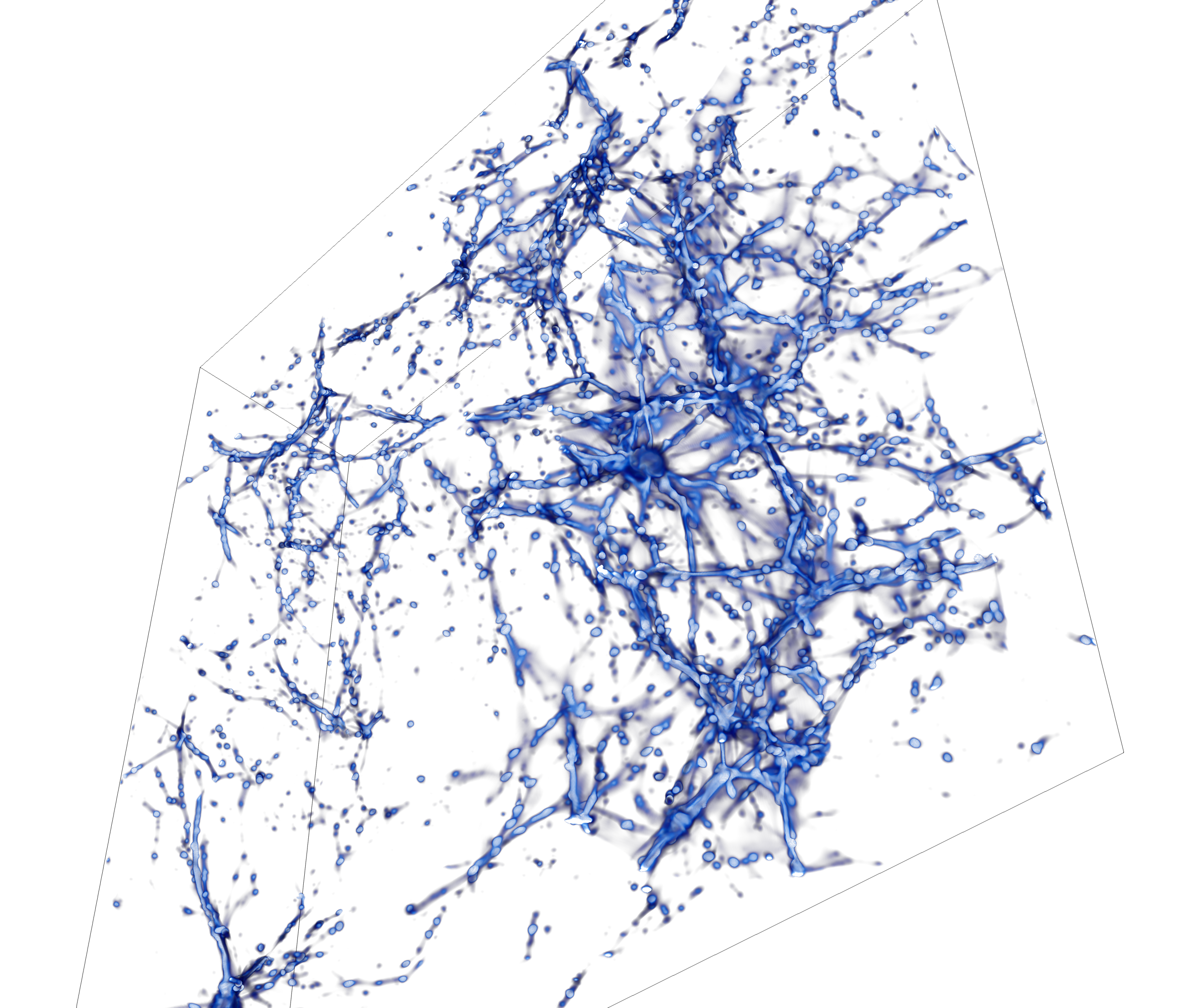}
        \caption[t]{Visualization of baryon density field of Nyx.}
       \vspace{-2mm}
        \label{fig:nyx-example}
\end{figure}

\begin{figure}[t]
     \centering
         \centering
         \vspace{-3mm}
         \includegraphics[width=0.7\columnwidth]{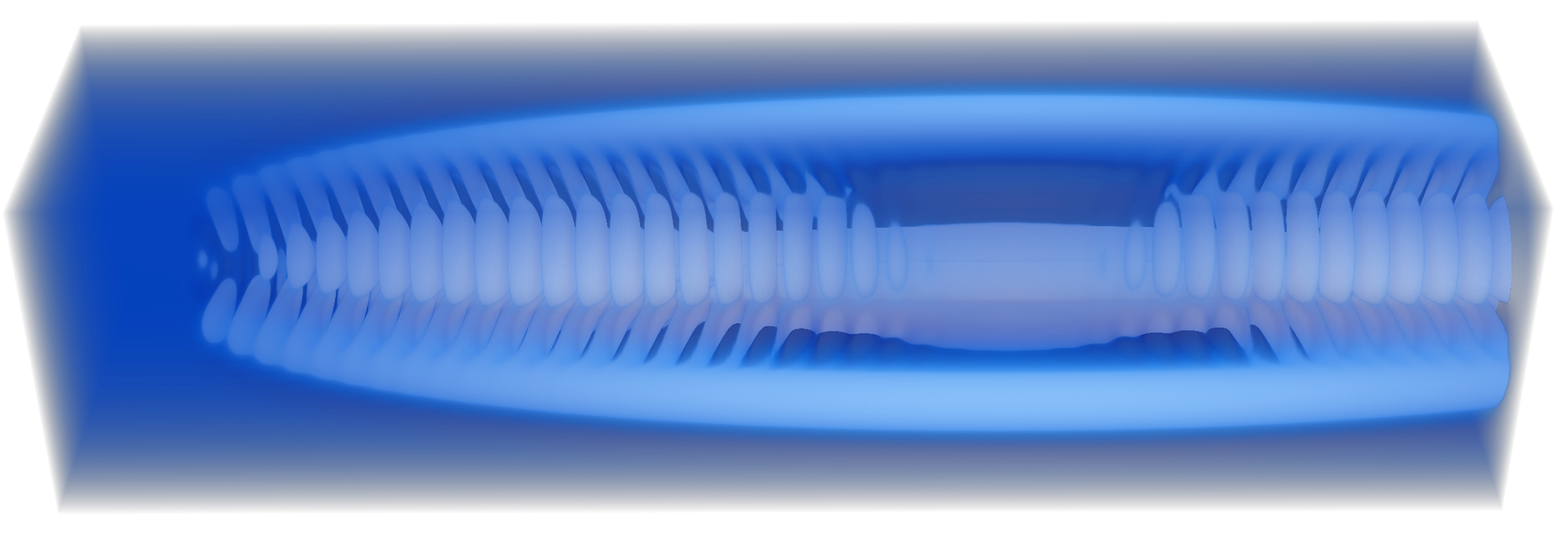}
        \caption[t]{Visualization of the electric field (x-direction) of WarpX.}
        \vspace{-2mm}
        \label{fig:wpx-example}
\end{figure}

\textbf{AMR applications.} Our evaluation primarily focuses on two AMR applications developed by the AMReX framework~\cite{zhang2019amrex}: Nyx cosmology simulation~\cite{nyx} and the WarpX~\cite{warpx} electromagnetic and electrostatic Particle-In-Cell (PIC) simulation. Nyx, as shown in Figure \ref{fig:nyx-example}, is a cutting-edge cosmology code that employs AMReX and combines compressible hydrodynamic equations on a grid with a particle representation of dark matter. Nyx generates six fields, including baryon density, dark matter density, temperature, and velocities in the $x$, $y$, and $z$ directions. WarpX, as shown in Figure \ref{fig:wpx-example}, is a highly-parallel and highly-optimized code that utilizes AMReX, runs on GPUs and multi-core CPUs, and features load-balancing capabilities. WarpX can scale up to the world's largest supercomputer and was the recipient of the 2022 ACM Gordon Bell Prize \cite{warpx-gordon}.

\textbf{Test platform.} Our test platform is the Summit supercomputer~\cite{summit} at Oak Ridge National Laboratory, each node of which is equipped with two IBM POWER9 processors with 42 physical cores and 512 GB DDR4 memory. It is connected to an IBM Spectrum Scale filesystem \cite{vef2016analyzing}. We use up to 128 nodes and 4096 CPU cores.

\textbf{Comparison baseline.} We compare our solution with AMReX's 1D SZ\_L/R compression solution \cite{amrexcomp} (denoted by ``AMReX''). We exclude zMesh and TAC because they are not in situ compression solutions, as mentioned in \S\ref{sec:introduction}. Note that we evaluate \thiswork{} with both SZ\_L/R (SZ with Lorenzo and linear regression predictors) and SZ\_Interp (SZ with spline interpolation predictor). 

\textbf{Test runs.}
As shown in Table~\ref{tab:dataset}, we have conducted six simulation runs in total, with three runs for each of the scientific applications, WarpX (WarpX\_1, WarpX\_2, and WarpX\_3) and Nyx (Nyx\_1, Nyx\_2, and Nyx\_3). Each simulation run consists of two levels, and the number of nodes (ranks) varying from 2 (64), to 16 (512), up to 128 (4096).
In WarpX\_1, the grid sizes of the levels progress from coarse to fine, with the dimensions of $256\times256\times2048$ and $512\times512\times4096$, and the data densities of 98.04\% and 1.96\%, respectively. The data size for one timestep in this run is 12.4 GB. For WarpX\_2, the grid sizes are $512\times512\times4096$ and $1024\times1024\times8192$, accompanied by the data densities of 98.05\% and 1.96\%. The data size for one timestep is 99.3 GB.
For WarpX\_3, the grid sizes are $1024\times1024\times8192$ and $2048\times2048\times16384$, accompanied by the data densities of 98.96\% and 1.04\%. The data size for one timestep is 624 GB.
Regarding Nyx\_1, the grid sizes of the levels, from coarse to fine, are $256\times256\times256$ and $512\times512\times512$, with data densities of 98.6\% and 1.4\%, respectively. The data size for one timestep in this run is 1.6 GB. For Nyx\_2, the grid sizes are $512\times512\times512$ and $1024\times1024\times1024$, and the data densities are 96.67\% and 3.23\%. The data size for one timestep is 12 GB. 
For Nyx\_3, the grid sizes are $1024\times1024\times1024$ and $2048\times2048\times2048$, and the data densities are 98.3\% and 1.7\%. The data size for one timestep is 97.5 GB.
Finally, as Summit uses a shared parallel filesystem that fluctuates in performance depending on the I/O load across the overall user-base,
we run each set of simulation runs multiple times and discard the results with abnormal performance (extremely slow).


\subsection{Evaluation on Compression Ratio} 
\label{eva:cr}

As demonstrated in Table~\ref{tab:cr}, our method, which includes optimized SZ\-L/R and optimized SZ\_Interp, outperforms the original 1D baseline for all three runs of the two applications, with a particularly notable improvement in WarpX. 
This superior performance is primarily due to our 3D compression's ability to leverage spatial and topological information, thereby enhancing the compression process. Moreover, the optimizations for SZ\_L/R and SZ\_Interp outlined in \S\ref{sec:design} further improve their respective compression performances.
Furthermore, as mentioned in \S\ref{sec:oph5}, the small chunk size of the original AMReX's compression leads to substantial encoding overhead, which ultimately results in a lower compression ratio.

Upon further analysis, we observe that WarpX exhibits a notably high compression ratio. This is mainly due to the smooth nature of the data generated by WarpX, as depicted in Figure~\ref{fig:wpx-example}, which results in excellent compressibility. In contrast, the data produced by Nyx appears irregular, as illustrated in Figure~\ref{fig:nyx-example}, making it more challenging to compress. Consequently, our \thiswork{} method has the potential to significantly reduce I/O time for WarpX by achieving greater data size reduction. Importantly, despite the difficulty in compressing Nyx data, our \thiswork{} will not introduce significant overhead to Nyx simulation, as will be demonstrated in \S\ref{eva:writetime}.

In terms of specific performance, SZ\-L/R proves more effective in handling Nyx data, while SZ\_Interp delivers superior compression ratios for WarpX. This can be attributed to SZ\-L/R's block-based predictor, which is better suited to capturing local patterns within Nyx data, whereas SZ\_Interp's global interpolation predictor excels when applied to the overall smoother data produced by WarpX.

\vspace{-2mm}
\begin{table}[h]
\small
\caption{Comparison of compression ratio (averaged across all fields/timesteps) with AMReX's original compression and \thiswork{}.}
\resizebox{.9\linewidth}{!}{
\begin{tabular}{|l|l|l|l|}
\hline
\textbf{Run}       & \textbf{AMReX(1D)} & \textbf{\thiswork(SZ\_L/R)} & \textbf{\thiswork(SZ\_Interp)}  \\ \hline
WarpX\_1 & 16.4  & 267.3        & \textbf{482.1} \\ \hline
WarpX\_2 & 117.5 & 461.2         & \textbf{2406.0}  \\ \hline
WarpX\_3 & 29.6 & 949.0         & \textbf{4753.7}  \\ \hline
Nyx\_1   & 8.8    & \textbf{15.0} & 14.0            \\ \hline
Nyx\_2   & 8.8    & \textbf{16.6} & 14.2            \\ \hline
Nyx\_3   & 8.7    & \textbf{16.3} & 13.6            \\ \hline
\end{tabular}}
\label{tab:cr}
\end{table}
\vspace{-6mm}

\begin{table}[b]
\caption{Comparison of reconstruction data quality (in PSNR) with AMReX's original compression and \thiswork{} for different runs.}
\small
\resizebox{.9\linewidth}{!}{
\begin{tabular}{|l|l|l|l|}
\hline
\textbf{Run}       & \textbf{AMReX(1D)} & \textbf{\thiswork(SZ\_L/R)} & \textbf{\thiswork(SZ\_Interp)}  \\ \hline
Nyx\_1   & 52.5   & \textbf{66.8} & 66.5           \\ \hline
Nyx\_2   & 56.7    & \textbf{69.1} & 68.9            \\ \hline
Nyx\_3   & 54.9    & \textbf{68.3} &  68.0            \\ \hline
WarpX\_1   & 73.6    & \textbf{80.3} & 79.9            \\ \hline
WarpX\_2   & 78.5   & 83.8 & \textbf{88.7}          \\ \hline
WarpX\_3   & 82.5   & 97.9 & \textbf{103.1}          \\ \hline
\end{tabular}}
\label{tab:psnr}
\end{table}

\subsection{Evaluation on Reconstruction Data Quality} 
\label{sec:psnr}
\begin{figure}[t]
     \centering
         \includegraphics[width=0.9\linewidth]{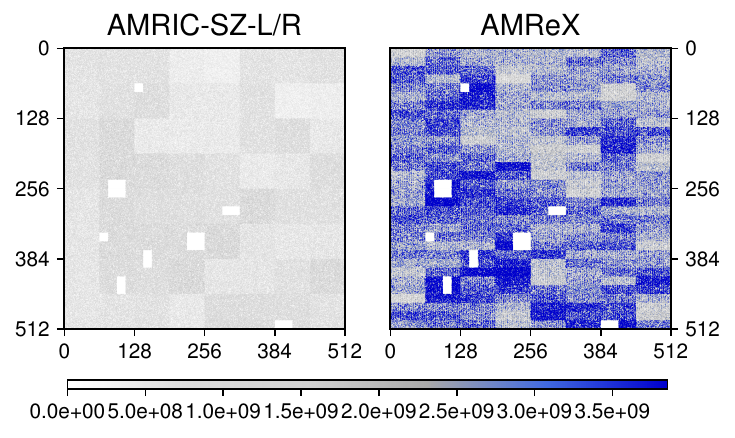}
        \caption[t]{Visualization comparison (one slice) of compression errors of our AMRIC (left) and AMReX's compression (right)  on Nyx\_2 ``baryon density'' field (i.e., coarse level, 96\% density, unit block size=32). Bluer/darker means higher compression error.}
        \label{fig:quality-vis}
\end{figure}

As shown in Table~\ref{tab:psnr}, the reconstruction data quality of \thiswork{} (including both SZ\_L/R and SZ\_Interp) is greatly higher than that of AMReX due to our optimization as well as the benefit of the 3D compression. As shown in Figure ~\ref{fig:quality-vis}, the error generated by AMRIC is considerably lower than that of AMReX. 
Note that the existing block-like pattern of the error is from the parallel compression done by 512 processes, each assigned with an error bound relative to the data range in the corresponding block. Therefore, the absolute error of each block is independent of those of the other blocks.

\begin{figure}[t]
     \centering
        \vspace{-5mm}
         \includegraphics[width=0.9\linewidth]{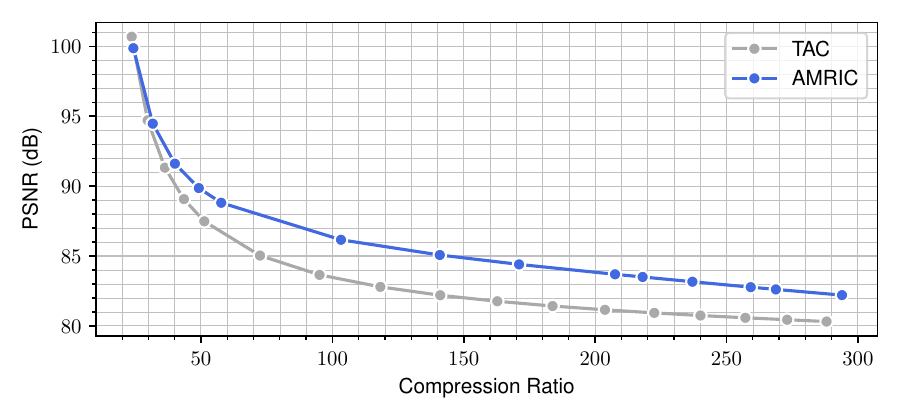}
         \vspace{-2mm}
        \caption[t]{Rate-distortion comparison of TAC and AMRIC using TAC's dataset \cite{tac-github} (i.e., Run1\_Z10).}
        \vspace{-2mm}
        \label{fig:vstac}
\end{figure}

\textit{\textbf{Comparison with offline solution TAC \cite{wang2022tac}:}}
To further demonstrate the effectiveness of AMRIC's optimized SZ\_L/R, we conduct a comparison with a state-of-the-art offline compression approach for 3D AMR data, called TAC (will be introduced in \S\ref{sec:related}), using the dataset from TAC's work.
As depicted in Figure~\ref{fig:vstac}, AMRIC outperforms TAC in terms of compression quality, achieving up to a 2.2$\times$ higher compression ratio while maintaining the same PSNR. This superior performance can be attributed to the fact that, unlike TAC, which only focuses on pre-processing and uses SZ\_L/R as a black box, AMRIC optimizes both the pre-processing and SZ\_L/R.

\textit{Insight of different compressors' performance on AMR simulations.} One takeaway is that, compared with SZ\_Interp, SZ\_L/R is more suitable for compressing AMR data because both AMR data and SZ\_L/R are block-based, while SZ\_Interp is global. Specifically, AMR simulations divide the data into boxes, which can negatively impact data locality and smoothness. SZ\_L/R, on the other hand, also truncates data into blocks, aligning with AMR simulations.  By applying our optimization approach outlined in \S\ref{sec:deco}, SZ\_L/R with SLE and adaptive block size can effectively mitigate the impact on data locality/smoothness caused by AMR applications, resulting in ideal compatibility with AMR applications.
On the other hand, since SZ\_Interp applies global interpolation to the unstructured block-based AMR data, it is challenging to achieve perfect compatibility between SZ\_Interp and AMR applications.

\subsection{Evaluation on I/O Time} 
\label{eva:writetime}
The overall writing time consists of: (1) pre-processing (including copying data to the HDF5 buffer, handling metadata, calculating the offset for each process) and I/O time without compression (directly writing the data to the disk); or (2) pre-processing and I/O time with compression (including compression computation cost and writing the compressed data to the file system). 

\begin{figure}[t]
     \centering
         \centering
         \includegraphics[width=\columnwidth]{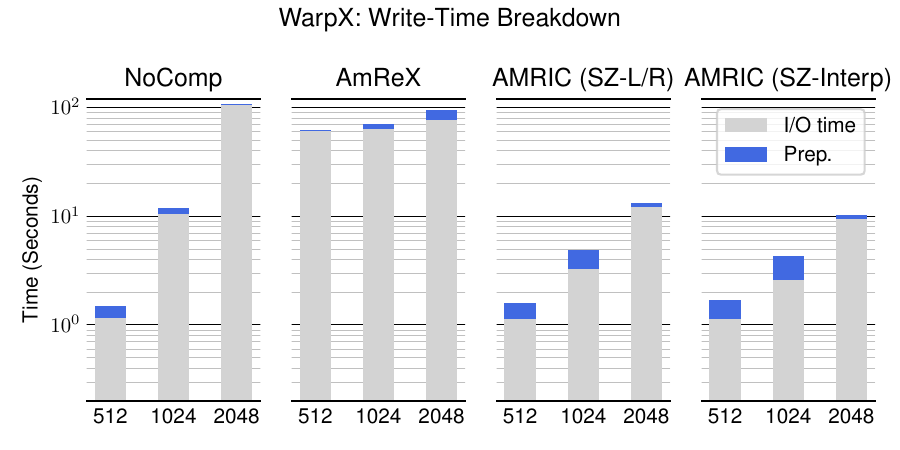}
         \vspace{-4mm}
        \caption[t]{Writing time of WarpX runs with different scales (in a weak scaling study). Log scale is used here for better comparison.}
        \label{fig:wpxtime}
        \vspace{-6mm}
\end{figure}

Figure~\ref{fig:wpxtime} shows that our in situ compression significantly reduces the overall writing time for HDF5 when compared to writing data without compression. This reduction reaches up to 90\% for the largest-scale WarpX run3, and 64\% for the larger-scale WarpX run2, without introducing any noticeable overhead for the smaller-scale WarpX run1.
This is because, for the relatively large-scale WarpX run3 and run2, the total data to be written amounts to 624 GB and 99 GB respectively. Due to the high compression ratio, we can significantly save both writing time and overall processing time.

\begin{figure}[t]
     \centering
         \centering
         \includegraphics[width=\columnwidth]{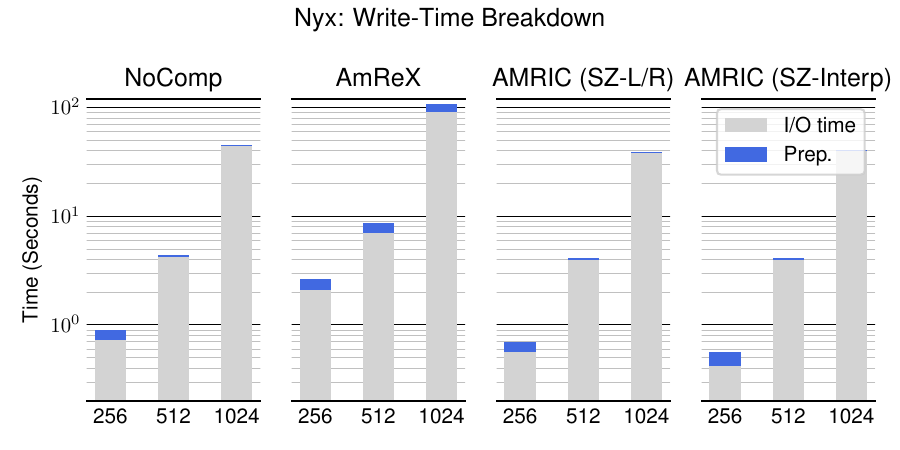}
         \vspace{-4mm}
        \caption[t]{Writing time of Nyx runs with different scales. Log scale is used for better comparison.}
       \vspace{-2mm}
        \label{fig:nyxtime}
\end{figure}

Our approach also considerably reduces the total writing time by 97\% compared to the original AMReX's compression for WarpX run1, 93\% for WarpX run2, and 89\% for WarpX run3. We can find out that AMReX's compression is extremely slow on WarpX. 
This is because, first, as discussed in \S\ref{sec:oph5} (Challenge 1), the original compression of AMReX is constrained by the existing data layout, which necessitates the use of a small HDF5 chunk size. This constraint negatively impacts compression time and ratio, causing suboptimal I/O performance. 
Moreover, this negative impact becomes more severe when there are relatively larger amounts of data in each process. Specifically, for all three WarpX runs, each process contains at least $128^3$ data points (considering only the coarse level). Consequently, given that the HDF5 chunk size for the original AMReX compression is 1024, each process will call the compressor 2048 times, generating a substantial startup cost and leading to extremely slow I/O. For WarpX run3, an HDF5 chunk size of 1024 causes issues, so we instead use 4096 as the chunk size.
This issue further emphasizes the significance of our modifications presented in \S\ref{sec:oph5}, which effectively utilize the HDF5 compression filter.
It is worth noting that the impact of this small chunk issue will be relatively mitigated when there are fewer data points in each process (as demonstrated in the subsequent evaluation of Nyx).

Also, note that our proposed method does not introduce any significant overhead to the pre-processing time.
This is primarily because our pre-processing strategy is lightweight, employing AMReX's built-in functions to identify redundant coarse data. Furthermore, the truncation, block ordering in the pre-processing stage, and data layout changes in \S\ref{sec:oph5} are performed simultaneously when loading data into the compression buffer, eliminating the need for extra rearrangement operations.
In addition, the elimination of redundant data using the pre-processing workflow reduces the amount of data to be processed, lowering the pre-processing time.

To further demonstrate the performance of our method on data with low compressibility as well as when each process owns fewer data, we conduct our test using Nyx with a smaller scale but the same number of nodes (ranks). In our Nyx runs, each process owns $64\times64\times64$ data points in the coarse level, which is 8 times less than that of WarpX runs (i.e., $128\times128\times128$). Note that a smaller data size in each process will lower the overall lossless encoding efficiency of the data, thus affecting the compression ratio.

As shown in Figure~\ref{fig:nyxtime}, even in a challenging setup (i.e., low data compressibility and low encoding efficiency), \thiswork{} is still able to achieve writing speeds comparable to those with no compression for all three Nyx runs. Furthermore, \thiswork{} significantly reduces the total writing time by 79\% compared to the original AMReX's compression for Nyx run1, by 53\% for Nyx run2, and by 64\% for Nyx run3.
It is worth noting that the small chunk issue in the Nyx run is relatively mitigated compared to WarpX. Specifically, we observe that the writing time using AMReX's compression for both 2-node and 16-node runs on Nyx is reduced by approximately 55 seconds, while the reduction for the 128-node run is approximately 10 seconds.
Our interpretation is that the time taken to launch the compression once remains constant (e.g., 0.03 seconds). In the Nyx run, each process needs to call the compressor only 256 times, as opposed to the 2048 calls required in the WarpX run1 and run2 (due to the increased HDF5 chunk size as aforementioned, WarpX run3 only requires 512 calls.). This difference results in a time reduction of $(2048-128)*0.03 \approx 55$ seconds for the writing process for WarpX run1 and run2, and $(512-128)*0.03 \approx 10$ for WarpX run3.

\vspace{-2mm}
\section{Related Work}
\label{sec:related}

There are two recent works focusing on designing efficient lossy compression methods for AMR datasets. Specifically, zMesh, proposed by Luo \textit{et al.} \cite{zMesh}, aims to leverage the data redundancy across different AMR levels. It reorders the AMR data across different refinement levels in a 1D array to improve the smoothness of the data. To achieve this, zMesh arranges adjacent data points together in the 1D array based on their physical coordinates in the original 2D/3D dataset. However, by compressing the data in a 1D array, zMesh cannot leverage higher-dimension compression, leading to a loss of topology information and data locality in higher-dimension data. On the other hand, TAC, proposed by Wang \textit{et al.} \cite{wang2022tac}, was designed to improve zMesh's compression quality through adaptive 3D compression. Specifically, TAC pre-processes AMR data before compression by adaptively partitioning and padding based on the data characteristics of different AMR levels. 

While zMesh and TAC provide offline compression solutions for AMR data, they are not designed for in situ compression of AMR data. In particular, zMesh requires extra communication to perform reordering in parallel scenarios. Specifically, zMesh must arrange neighboring coarse and fine data more closely. However, the neighboring fine and coarse data might not be owned by the same MPI rank. As a result, data from different levels must be transferred to the appropriate processes, leading to high communication overhead.
TAC requires the reconstruction of the entire physical domain's hierarchy to execute its pre-processing approach. This relatively complex process results in significant overhead for in situ data compression.

Although the AMReX framework \cite{zhang2019amrex} supports in situ AMR data compression through HDF5 compression filters \cite{amrexcomp}, it has two main drawbacks:
(1) The original AMReX compression only compresses the data in 1D, limiting its capacity to benefit from higher-dimension compression and resulting in sub-optimal compression performance, particularly in terms of compression quality.
(2) The original AMReX compression cannot effectively utilize the HDF5 filter. Specifically, due to the limitation of the data layout for multiple fields, AMReX can only adopt a very small chunk size to prevent compressing different physical fields together. As a result, AMReX needs to apply the compressor separately for each small chunk, resulting in low I/O and compression performance as demonstrated in \S\ref{eva:writetime}, \S\ref{eva:cr}, and \S\ref{sec:psnr}.

\section{Conclusion and Future Work}
\label{sec:conclusion}
In conclusion, we have presented \thiswork{}, an effective in situ lossy compression framework for AMR simulations that significantly enhances I/O performance and compression quality. Our primary contributions include designing a compression-oriented pre-processing in situ workflow for AMR data, optimizing the state-of-the-art SZ lossy compressor for AMR data, efficiently utilizing the HDF5 compression filter on AMR data, and integrating \thiswork{} into the AMReX framework. We evaluated \thiswork{} on two real-world AMReX applications, WarpX and Nyx, using 4096 CPU cores from the Summit supercomputer. The experimental results demonstrate that \thiswork{} achieves up to 10.5$\times$ I/O performance improvement over the non-compression solution and up to 39$\times$ I/O performance improvement and up to 81$\times$ compression ratio improvement with better data quality over the original AMReX's compression solution. 

In future work, we plan to evaluate \thiswork{} on additional AMReX applications accelerated by GPUs. Furthermore, we will assess \thiswork{} on a wider range of HPC systems and at different scales. 
Additionally, we will incorporate our in situ compression solution into other AMR frameworks.


\section*{Acknowledgement}
\small
This work (LA-UR-23-24096) has been authored by employees of Triad National Security, LLC which operates Los Alamos National Laboratory under Contract No. 89233218CNA000001 with the U.S. Department of Energy and National Nuclear Security Administration. 
The material was supported by the U.S. Department of Energy, Office of Science and Office of Advanced Scientific Computing Research (ASCR), under contract DE-AC02-06CH11357. 
This work was partly supported by the Exasky Exascale Computing Project (17-SC-20-SC), a collaborative effort of the U.S. Department of Energy Office of Science and the National Nuclear Security Administration. 
This work was also supported by NSF Grants OAC-2003709, OAC-2303064, OAC-2104023, OAC-2247080, 
OAC-2311875, OAC-2311876, OAC-2312673.

\newpage
\bibliographystyle{ACM-Reference-Format}
\bibliography{refs}

\end{document}